\documentclass[aps,prx,reprint,groupedaddress]{revtex4-1}
\usepackage{url}
\usepackage[dvips]{graphicx}
\usepackage{amsthm}
\usepackage{amsmath}
\usepackage{physics}
\usepackage{here}
\usepackage{amssymb}
\usepackage{color}
\usepackage{dcolumn}
\usepackage{bm}
\usepackage{bbold}
\usepackage{letltxmacro}
\usepackage[usenames,dvipsnames]{xcolor}
\usepackage[breaklinks,colorlinks=true,citecolor=blue,linkcolor=RubineRed,urlcolor=RubineRed]{hyperref}
\NewDocumentCommand{\evalat}{sO{\big}mm}{%
  \IfBooleanTF{#1}
   {\mleft. #3 \mright|_{#4}}
   {#3#2|_{#4}}%
}

\newtheorem{theorem}{Theorem}
\newtheorem{lemma}{Lemma}
\newtheorem{proposition}{Proposition}

\newtheorem{corollary}{Corollary}
\newcommand{\coc}[1]{\relax{#1}}
\newcommand{\cod}[1]{\relax{#1}}

\newcommand{\com}[1]{\relax{#1}}
\newcommand{\coa}[1]{\relax{#1}}
\newcommand{\cob}[1]{\relax{#1}}

\allowdisplaybreaks[0]

\begin{document}
\title{Universal noise-precision relations in variational quantum algorithms}

\author{Kosuke Ito$^{1}$}\email{kosuke.ito@qc.ee.es.osaka-u.ac.jp} 
\author{Wataru Mizukami$^{1, 2, 3}$}\email{wataru.mizukami.857@qiqb.osaka-u.ac.jp}
\author{Keisuke Fujii$^{1, 2, 4}$}\email{fujii@qc.ee.es.osaka-u.ac.jp}
\affiliation{
${}^1$Center for Quantum Information and Quantum Biology, International Advanced Research Institute, Osaka University, Osaka 560-8531, Japan\\
${}^2$Graduate School of Engineering Science, Osaka University, 1-3 Machikaneyama, Toyonaka, Osaka 560-8531, Japan\\
${}^3$JST, PRESTO, 4-1-8 Honcho, Kawaguchi, Saitama 332-0012, Japan\\
${}^4$RIKEN Center for Quantum Computing (RQC),
Hirosawa 2-1, Wako, Saitama 351-0198, Japan}

\begin{abstract}
 Variational quantum algorithms (VQAs) are expected to become a practical application of near-term noisy quantum computers. Although the effect of the noise crucially determines whether a VQA works or not, the heuristic nature of VQAs makes it difficult to establish analytic theories. Analytic estimations of the impact of the noise are urgent for searching for quantum advantages, as numerical simulations of noisy quantum computers on classical computers are heavy and quite limited to small scale problems. In this paper, we establish analytic estimations of the error in the cost function of VQAs due to the noise. The estimations are applicable to any typical VQAs under Gaussian noise, which is equivalent to a class of stochastic noise models. Notably, depolarizing noise is included in this model. As a result, we obtain estimations of the noise level to guarantee a required precision. Our formulae show how the Hessian of the cost function, the spectrum of the target operator, and the geometry of the ansatz affect the sensitivity to the noise. This insight implies trade-off relations between the trainability and the noise resilience of the cost function. We also obtain rough estimations which can be easily calculated without detailed information of the cost function. As a highlight of the applications of the formula, we propose a quantum error mitigation method which is different from the extrapolation and the probabilistic error cancellation.
\end{abstract}

\maketitle

\section{Introduction}
To make use of noisy intermediate-scale quantum (NISQ) devices in the near future \cite{Preskill2018quantumcomputingin},
we have to seek a classically intractable task that hundreds of qubits can resolve under the lack of the error correction.
A promising framework to realize it is hybrid quantum-classical algorithms, where most of the processes are done on a classical computer, receiving the output from a quantum circuit which computes some classically intractable functions.
Especially, variational quantum algorithms (VQAs) have attracted much attention, where the cost function of a variational problem is computed by utilizing low-depth quantum circuits and the optimization of the variational parameters is done on a classical computer.
For example, the variational quantum eigensolver (VQE) \cite{Peruzzo:2014uz,PhysRevX.6.031045,Kandala:2017wb} is a VQA to obtain an approximation of the ground state of a Hamiltonian, and beyond \cite{McClean2017,Santagatieaap9646,1810.12745,PhysRevX.8.011021,McArdle:2019we,PhysRevA.99.062304,Parrish2019,Higgott2019,PhysRevResearch.1.033062,tilly2020computation,Ollitrault2019}. The quantum approximate optimization algorithm (QAOA) \cite{1411.4028,1602.07674,1712.05771} is another attracting VQA for combinatorial optimization problems.
Quantum machine learning algorithms \cite{Biamonte:2017ug,Benedetti_2019} for NISQ devices have also been proposed in various settings \cite{1307.0411,1412.3489,PhysRevA.98.012324,PhysRevA.98.032309,Khoshaman_2018,Havlicek:2019tu}.

The noise is one of the most crucial obstacles to overcome toward achieving quantum advantage via VQAs.
The heuristic nature of VQAs makes it difficult to analytically assess the effects of the noise on the performance of VQAs.
To go beyond heavy numerical simulations of noisy quantum computers on classical computers limited to small scale problems, analytic estimations of the impact of the noise are urgent for obtaining knowledge about intermediate scale problems with potential quantum advantage.
In fact, this issue has been actively studied in recent years, and some analytic results have been obtained, for example,
on the characterization of the impact of local noise in QAOA \cite{Marshall_2020},
noise resilience of the optimization results \cite{Sharma_2020, PhysRevA.104.022403}, noise-induced barren plateaus \cite{Wang:2021wb}, noise-induced breaking of symmetries \cite{Fontana2022nontrivial}, effects of the noise on the convergence property of the optimizations in VQAs \cite{gentini_noise-resilient_2020}.

In this work, we establish analytic estimation formulae on the error in the cost function of VQAs due to the noise, which are applicable to any typical VQAs under Gaussian noise.
Especially, we focus on the effect of the noise on the expectation value in order to investigate ultimately achievable and unachievable precision, aside from the statistical error due to the finiteness of the number of measurements.
\coc{Gaussian noise is equivalent to a class of stochastic noise models given in Eq.~(\ref{isuelzgp}). Notably, depolarizing noise can be decomposed into this form of the stochastic noise channels, and hence, is included in this model.}
The correspondence from the stochastic noise model to the Gaussian model is given by introducing virtual parametric gates associated with the noise.
Our formulae essentially come from the expansion of the cost function with respect to the fluctuations in the parameters due to the noise.
This fact implies that a picture of the noise based on fluctuations of parameters of virtual parametric gates can serve as a powerful tool for performance analysis of VQAs.
In fact, we propose a quantum error mitigation method based on this expansion including the virtual parameters, which is different from existing error mitigation methods such as the extrapolation \cite{PhysRevX.7.021050,PhysRevLett.119.180509} and the probabilistic error cancellation \cite{PhysRevLett.119.180509, PhysRevX.8.031027}.

\coc{Applying our formulae, we can estimate the order of magnitude of both sufficient noise level and the necessary one to achieve a desired precision.
Moreover, we can gain an insight of what properties of the problem affect the sensitivity of the cost function to the noise.
More concretely, our formulae implies that the sensitivity to the noise is affected by the Hessian of the cost function, or the spectrum of the target operator and the geometry of the ansatz.
Trade-off relations between the trainability and the noise resilience of the cost function are repeatedly implied in some forms as a result of our formulae.}

We also obtain computable rough upper and lower bounds of the precision of the noisy cost function for a VQA task, whose usefulness is verified in numerical simulations of \coc{the Heisenberg spin chain} and a toy model.

The rest of the paper is organized as follows.
In Sec.~\ref{jvbilzhd}, we describe the setup of the VQA under the Gaussian noise.
The correspondence between the Gaussian and the stochastic noise models including depolarizing noise model is shown in Sec.~\ref{ssec-correspond}.
In Sec.~\ref{jvxylzhd}, the main theorem (Theorem \ref{thm1}) is shown.
Then, estimations of a sufficient order of the smallness of the noise to achieve a given precision is followed.
In Sec.~\ref{jshqlzhb}, we propose an error mitigation method based on the main theorem.
Next, in Sec.~\ref{jvfalzhb} we establish upper and lower bounds of the error in the cost function, which show how the spectrum of the target operator and the geometric structure of the ansatz affect the sensitivity of the cost function to the noise.
An estimation of a necessary order of the smallness of the noise to achieve a required precision is followed.
We provide rough estimations which can be easily calculated without detailed information of the cost function at the tail of Sec.~\ref{jvfalzhb}.
In Sec.~\ref{sec-numerics}, we demonstrate the usefulness of the rough estimations by numerical simulations of Heisenberg spin chain and a toy model.
The conclusion is drawn in Sec.~\ref{jzbmlzhd}.
\cod{A summary table of important notations is presented in Table.~\ref{tab:symbols_parameters} in Appendix.~\ref{App_notations} for convenience.}

\section{Setup}
\label{juxulzhd}
\subsection{Gaussian noise model of the parameterized quantum circuit}
\label{jvbilzhd}
We consider the following parameterized quantum circuit
\begin{align}
 U(\vec{\theta}) = \prod_{i=1}^{M} U_i(\theta_i) W_i,
\end{align}
where $U_i(\theta_i) = \exp [-i \theta_i A_i / 2]$ satisfying $A_i^2 = I$ with the identity operator $I$, and $W_i$ is a generic non-parametric gate.
\com{Typical parameterized quantum circuits such as the hardware efficient ansatz \cite{Kandala:2017wb,Havlicek:2019tu} satisfy the above requirements.}
\com{We focus on a VQA to minimize
the cost function $C(\vec{\theta})$ given by the sum of the expectation values of the target Hermitian operators $H_l$ $(l=1,2,\cdots,L)$ as
\begin{align}
 C(\vec{\theta}) = \sum_{l=1}^{L}\bra{\phi_l} U(\vec{\theta})^{\dagger} H_l U(\vec{\theta}) \ket*{\phi_l},
\end{align}
where $\ket*{\phi_l}$ $(l = 1,2,\cdots, L)$ are the input states.
}
\coc{As a model of the noise, we consider independent Gaussian noise in the parameter, where each parameter $\theta_i$ independently fluctuates as $\theta_i + \eta_i$ with the Gaussian random variable $\eta_i$ with zero mean and the variance $\sigma_i^2$. In other words, the Gaussian noise channel $\mathcal{G}_{A_i,\sigma_i}$ defined below is inserted after each $U_i(\theta_i)$:
\begin{align}
 \mathcal{G}_{A_i,\sigma_i}(\rho) :=& \int_{-\infty}^{\infty} \mathcal{U}_{A_i,\eta}(\rho) \frac{e^{-\frac{\eta^2}{2\sigma_i^2}}}{\sqrt{2\pi}\sigma_i} d\eta\nonumber\\
=&\int_{-\infty}^{\infty} \mathcal{U}_{A_i,\eta}(\rho) f_{\sigma_i}(\eta) d\eta,\label{Gauss_ch_def}
\end{align}
where $f_{\sigma} = e^{-\frac{\eta^2}{2\sigma^2}}/(\sqrt{2\pi}\sigma)$ is the probability density function of the zero-mean Gaussian distribution with the variance $\sigma^2$, $\rho$ is any density operator, and $\mathcal{U}_{A_i,\eta}(\rho):= e^{-i\frac{\eta}{2}A_i}\rho e^{i\frac{\eta}{2}A_i}$.}
\coc{The incompleteness of the control and statistical error in the parameters obtained as a result of an optimization (e.~g.~in the stochastic gradient descent \cite{Sweke2020stochasticgradient}) may result in such fluctuations in the parameters.
Moreover, considering Gaussian noise in ``virtual parameters'', we can also treat stochastic noise models as the Gaussian noise model as shown in the next section.}

In this paper, we only focus on the effect of the noise on the expectation value in order to investigate ultimately achievable and unachievable precision, aside from the statistical error due to the finiteness of the number of measurements.

\subsection{Correspondence to the stochastic noise model}\label{ssec-correspond}

Here, we show the correspondence relation between the Gaussian and the stochastic noise models along the same lines \com{with Nielsen and Chuang's textbook} \cite{nielsen2010quantum}.
We consider the case where $M_{\mathrm{SNC}}$ stochastic noise channels
\begin{align}
 \mathcal{E}_{B_{\nu},p_{\nu}}(\rho) := (1 - p_{\nu}) \rho + p_{\nu} B_{\nu} \rho B_{\nu}\label{isuelzgp}
\end{align}
with respect to operators $B_\nu$ $(\nu=1,2,\cdots, M_{\mathrm{SNC}})$, $B_{\nu}^2=I$ are inserted in the circuit, where $\rho$ denotes a density operator, $0<p_{\nu}<1/2$ is the error probability.
Then we have the following correspondence between the Gaussian and the stochastic noise models:
\cod{
\begin{proposition}\label{prop1}
The relation
 \begin{align}
\mathcal{G}_{B_\nu,\sigma_{\mathrm{SNC}, \nu}}(\rho)
= \mathcal{E}_{B_\nu,p_\nu}(\rho)\label{itwtlzgw}
\end{align}
holds with the corresponding variance
\begin{align}
 \sigma_{\mathrm{SNC}, \nu}^2 = - 2 \log (1 - 2 p_\nu).\label{Gauss-stoch}
\end{align}
\end{proposition}}

Hence, if we consider the stochastic $B_\nu$-noise $(\nu=1,2,\cdots,M_{\mathrm{SNC}})$, it
can be treated as Gaussian noise with respect to the virtually inserted parametric gate $V_\nu(\xi_\nu) := \exp [-i \xi_\nu B_\nu / 2]$ at the place where the noise occurs, where $\xi_\nu \equiv 0$ throughout the optimization.
Therefore, the cost function $C_{\mathrm{noisy}}(\vec{\theta})$ evaluated under the stochastic noises and the fluctuations in the optimizing parameters
is given as
\begin{align}
 &C_{\mathrm{noisy}}(\vec{\theta})\nonumber\\
=& \int C(\vec{\theta} + \vec{\eta}, \vec{\Delta}) \prod_{j=1}^{M} f_{\sigma_{j}}(\eta_j) d\eta_j \prod_{\nu=1}^{M_{\mathrm{SNC}}} f_{\sigma_{\mathrm{SNC},\nu}}(\Delta_\nu) d\Delta_\nu.\label{Cnoisy}
\end{align}
In the following, $C(\vec{\theta})$ denotes the abbreviation of $C(\vec{\theta},\vec{\xi}) = C(\vec{\theta}, \vec{0})$.
Especially, the partial derivative of the cost function with respect to a virtual parameter $\xi_\nu$ at $(\vec{\theta},\vec{\xi})=(\vec{\theta},\vec{0})$ is denoted by $\frac{\partial}{\partial \xi_\nu} C(\vec{\theta})$.
\coc{Hereafter, $\nu$ is used only to denote the indices of the stochastic noise channels and its corresponding virtual parameters in distinction from those of the optimized parametric gates.
A benefit of introducing the virtual parameters is to treat stochastic noises mathematically in the same way as the fluctuations in the parameters.
However, it should be noted that the virtual parameters $\vec{\xi}$ are just fixed to zero and nothing to do with the optimization.
We call $\theta_j$ a optimizing parameter in distinction from a virtual parameter.
Nevertheless, 
there are cases where a virtual parameter $\xi_\nu$ is equivalent to an optimizing parameter $\theta_j$ when $B_{\nu}$-stochastic noise occurs alongside the parametric gate $U_j(\theta_j)$ with $A_j=B_{\nu}$.
One example is when the generator of the parametric gate is a Pauli operator and a depolarizing channel is applied after this gate. In this case, the depolarizing channel is decomposed into stochastic noise channels with respect to all the Pauli operators, so one of the virtual parameters $\xi_\nu$ coincides with the optimizing parameter $\theta_j$ for the Pauli operator used in the gate.
Especially, the relation $\frac{\partial}{\partial \xi_\nu} = \frac{\partial}{\partial \theta_j}$ gives a connection between the trainability of the optimizing parameters and the sensitivity to the stochastic noise as seen later.}
We also remark that a similar correspondence to Eq.~(\ref{itwtlzgw}) holds not only for Gaussian noise but also for any noise in the parameter whenever its probability density function is even, since only this property is used to show Eq.~(\ref{itwtlzgw}).

\com{Especially, depolarizing noise is one of the most basic and serious error sources for noisy quantum computers.
A key feature of the Gaussian noise model is its capability of treating depolarizing noise via the above correspondence.}
Depolarizing noise is described by the depolarizing channel
\begin{align}
 \mathcal{D}_{k,q}(\rho) = (1 - p) \rho + p(4^k - 1)^{-1} \sum_{i=1}^{4^k - 1} P_i \rho P_i,\label{DP_def}
\end{align}
where $P_i$ runs over all $k$-qubit Pauli operators except for the identity $I =: P_0$, and $p$ is the error probability.
\com{Since we can decompose the depolarizing channel into multiple stochastic noise channels with respect to each single Pauli operator, the above correspondence works.}
\cod{\begin{lemma}\label{Lem-gauss-dep}
Let $p < (4^k - 1) / 4^k$.
 The $k$-qubit depolarizing channel $\mathcal{D}_{k,p}$ can be decomposed as $\mathcal{D}_{k,p} = \prod_{i=1}^{4^k - 1} [(1-\tilde{p})\mathcal{I} + \tilde{p}\mathcal{U}_{P_i}]$ into $4^k - 1$ stochastic Pauli noise channels with respect to $k$-qubit Pauli operators $P_i$, where $\mathcal{I}$ is the identity channel, $\mathcal{U}_{P_i}(\rho) = P_i \rho P_i$ for arbitrary state $\rho$, and the corresponding error probability $\tilde{p}$ is given as
\begin{align}
 \tilde{p} = \frac{1}{2}\left[1 - \left(1-\frac{4^k}{4^k - 1}p\right)^{\frac{1}{2\cdot 4^{k-1}}}\right].\label{eq-pauli-d}
\end{align}
Equivalently, the $k$-qubit depolarizing channel $\mathcal{D}_{k,p}$ can be decomposed into the Gaussian noise channels as
$\mathcal{D}_{k,p} = \prod_{i=1}^{4^k - 1} \mathcal{G}_{P_i,\sigma_{\mathrm{DP}}(k)}$
with the common variance
\begin{align}
 \sigma_{\mathrm{DP}}^2(k) = - \frac{1}{4^{k-1}} \log \left(1-\frac{4^k}{4^k - 1}p\right).\label{eq-gauss-d}
\end{align}
\end{lemma}}

A proof of Lemma \ref{Lem-gauss-dep} is in Appendix \ref{proof_lem1}.
We remark that Eq.~(\ref{eq-gauss-d}) implies that
\begin{align*}
 \sigma_{\mathrm{DP}}^2(k) = \frac{4}{4^k - 1}p + O\left(p^2\right)
\end{align*}
for small error probability $p$ in the same way as Eq.~(\ref{sgm-sim-p}).

\section{Universal Error Estimations}
\subsection{Estimation of the leading-order term of the error}
\label{jvxylzhd}
\coc{At first, we show an estimation of the error in the cost function due to the fluctuations in the parameters following a general probability measure not restricted to the Gaussian distribution. Because the fluctuations in the virtual parameters associated with the stochastic noises can be treated totally in the same way as those in the optimizing parameters,
we treat the optimizing and virtual parameters together in the same notation as
$\theta_{M + \nu} = \xi_\nu$ $(\nu = 1, 2, \cdots, M_{\mathrm{SNC}})$, $M_{\mathrm{tot}} := M + M_{\mathrm{SNC}}$, and $\sigma_{M + \nu} := \sigma_{\mathrm{SNC},\nu}$.
We state the following main theorems in the above notation for brevity.
We will use the same notation also in the subsequent sections when it is convenient to treat the optimizing and virtual parameters together in the same manner.
\begin{theorem}\label{thm0}
 Let us assume that each parameter independently fluctuates as $\theta_i + \eta_i$, where $\eta_i$ is a zero-mean random variable with probability measure $\mathcal{P}_i$.
We assume that the moment generating function (mgf) $g_i(t) = \int \exp(\eta_i t) d\mathcal{P}_i(\eta_i)$ of each $\eta_i$ exists and is analytic in a region including $0$ and $1$.
We also assume that every odd moment are nonnegative: $\int \eta_i^{2\alpha + 1} d \mathcal{P}_i(\eta_i) \geq 0$, where $\alpha$ is any positive integer.
Then, the noisy cost function
\begin{align}
 \tilde{C}(\vec{\theta}) = \int C(\vec{\theta} + \vec{\eta})\prod_{i} d\mathcal{P}_i(\eta_i)\label{gen_Cnoisy}
\end{align}
with respect to this noise is estimated as follows:
\begin{align}
 &\left|\tilde{C}(\vec{\theta}) - C(\vec{\theta}) - \frac{1}{2} \sum_{i=1}^{M_{\mathrm{tot}}} \frac{\partial^2}{\partial \theta_i^2} C(\vec{\theta}) \sigma_i^2\right|\nonumber\\
\leq& \frac{\sum_{l=1}^{L}(E_{\max,l} - E_{0,l})}{2} \left[\prod_{i=1}^{M_{\mathrm{tot}}} g_i(1) - \frac{1}{2}\sum_{i=1}^{M_{\mathrm{tot}}} \sigma_i^{2} - 1 \right],\label{thm0ineq}
\end{align}
\cod{where $E_{0,l}$, $E_{\max,l}$ are the minimum and the largest eigenvalues of $H_l$, respectively,} and
$\sigma_i^2$ is the variance of $\eta_i$.
\end{theorem}
A proof of Theorem \ref{thm0} is given in Appendix \ref{thm0_proof}. Theorem \ref{thm0} gives a bound of the precision of an approximation of the error $\tilde{C}(\vec{\theta}) - C(\vec{\theta})$ by $\frac{1}{2} \sum_{i=1}^{M_{\mathrm{tot}}} \frac{\partial^2}{\partial \theta_i^2} C(\vec{\theta}) \sigma_i^2$.
We remark that the Taylor expansion of the mgf reads
\begin{align}
 g_i(1) = \sum_{k=0}^{\infty} \frac{\mu_i^{(k)}}{k!},\label{taylor-mgf}
\end{align}
where $\mu_i^{(k)}:= \int \eta_i^k d\mathcal{P}_i(\eta_i)$ is the $k$-th moment of $\mathcal{P}_i$. Especially, $\mu_i^{(0)} = 1$, $\mu_i^{(1)} = 0$ and $\mu_i^{(2)} = \sigma_i^2$.
Thus, the terms $1 + \frac{1}{2}\sum_{i=1}^{M_{\mathrm{tot}}} \sigma_i^{2}$ are canceled out inside the bracket of the right hand side of Eq.~(\ref{thm0ineq}).
Hence, whether this approximation is effective or not depends on the behavior of third and higher-order moments of $\mathcal{P}_i$.
We leave for future work the detailed analysis of noise with general probability distributions. In the following, we return to focusing on Gaussian noise.}

\coc{We can apply Theorem \ref{thm0} to the Gaussian distribution because every odd moment of the zero-mean Gaussian distribution is zero, and its mgf $\exp(\sigma_i^2 t^2/2)$ obviously satisfies the assumptions of Theorem \ref{thm0}.
Therefore, we obtain the following leading-order approximation of the error \com{$\epsilon(\vec{\theta}) := C_{\mathrm{noisy}}(\vec{\theta}) - C(\vec{\theta})$} in the cost function due to Gaussian noise from Theorem \ref{thm0}:}
\begin{theorem}\label{thm1}
We have the following estimation of the deviation of the cost function due to the fluctuations in the parameters following the Gaussian distribution:
 \begin{align}
 &
\left|\epsilon(\vec{\theta}) - \frac{1}{2} \sum_{i=1}^{M_{\mathrm{tot}}} \frac{\partial^2}{\partial \theta_i^2} C(\vec{\theta}) \sigma_i^2
\right|
\nonumber\\
\leq
& 
\frac{\sum_{l=1}^{L}(E_{\max,l} - E_{0,l})}{2} \left[\exp(\frac{1}{2}\sum_{i=1}^{M_{\mathrm{tot}}} \sigma_i^{2}) - \frac{1}{2}\sum_{i=1}^{M_{\mathrm{tot}}} \sigma_i^{2} - 1 \right]\nonumber\\
=& \sum_{l=1}^{L}(E_{\max,l} - E_{0,l})\left[\frac{1}{16}\left(\sum_{i=1}^{M_{\mathrm{tot}}} \sigma_i^{2}\right)^2 + O\left(\left(\sum_{i=1}^{M_{\mathrm{tot}}} \sigma_i^{2}\right)^3\right)\right]\nonumber\\
=& O\left(\sum_{l=1}^{L}(E_{\max,l} - E_{0,l}) \left(\sum_{i=1}^{M_{\mathrm{tot}}} \sigma_i^{2}\right)^2 \right).\label{eq-thm1}
 \end{align}
\end{theorem}

\coc{We remark that a similar analysis to Theorem \ref{thm1} appears in Ref.\cite{2212.09431}.}
Theorem \ref{thm1} implies that the error $\epsilon(\vec{\theta})$ is approximated as
\begin{align}
\epsilon(\vec{\theta}) \approx
 \frac{1}{2} \sum_{i=1}^{M_{\mathrm{tot}}} \frac{\partial^2}{\partial \theta_i^2} C(\vec{\theta}) \sigma_i^2\label{berib}
\end{align}
if the variances $\sigma_i^2$ are small enough so that $\sum_{l=1}^{L}(E_{\max,l} - E_{0,l})\left(\sum_{i=1}^{M_{\mathrm{tot}}} \sigma_i^{2}\right)^2$ is sufficiently small.
For typical problems, $\sum_{l=1}^{L}(E_{\max,l} - E_{0,l}) \leq 2\sum_{l=1}^{L}\|H_l\|$ is in polynomial order of the number of qubit $n$, i.~e.,~$\sum_{l=1}^{L}(E_{\max,l} - E_{0,l}) = O(n^r)$ with a positive number $r$ (e.~g.~$r = 1$ for locally interacting spin systems, $r=4$ for the Jordan-Wigner transformed full configuration interaction Hamiltonian of molecules \cite{1928ZPhy47631J,helgaker2014molecular,Peruzzo:2014uz}).
Then, if all the variances are in the same order $\sigma_i^2= O(\sigma^2)$ $(i=1,2,\cdots, M_{\mathrm{tot}})$,
this approximation is valid when $\sigma^2 = o\left(n^{-\frac{r}{2}}{M_{\mathrm{tot}}}^{-1}\right)$ in the sense that $\sum_{l=1}^{L}(E_{\max,l} - E_{0,l}) \left(\sum_{i=1}^{M_{\mathrm{tot}}} \sigma_i^{2}\right)^2 = o(1)$.

\coc{Now, we explicitly apply Theorem \ref{thm1} to the virtual parameters associated with the stochastic noises, and rewrite the estimation in terms of the error probability.
Eq.~(\ref{Gauss-stoch}) implies that
\begin{align}
 \sigma_{\mathrm{SNC}, \nu}^2 = 4 p_\nu + O\left(p_\nu^2\right)\label{sgm-sim-p}
\end{align}
for small error probability $p_\nu$ from the Taylor expansion $-\log (1-x) = x + O(x^2)$.
Eq.~(\ref{Gauss-stoch}) implies that
\begin{align}
 \sigma_{\mathrm{SNC}, \nu}^2 = 4 p_\nu + O\left(p_\nu^2\right)\label{sgm-sim-p}
\end{align}
from the Taylor expansion $-\log (1-x) = x + O(x^2)$.}
\coc{Then, applying Theorem \ref{thm1} to the virtual parameters with the relations (\ref{itwtlzgw}) and \eqref{sgm-sim-p}, we obtain the following corollary:
\begin{corollary}\label{cor1}
 Let the stochastic noise channels $\mathcal{E}_{B_\nu,p_\nu}(\rho) = (1-p_\nu)\rho + p_\nu B_\nu\rho B_\nu$ $(\nu = 1, 2, \cdots, M_{\mathrm{SNC}})$ with the error probability $0<p_\nu<1/2$ be inserted in the circuit with fluctuating optimizing parameters due to Gaussian noise, and hence the noisy cost function is given as Eq.~(\ref{Cnoisy}).
Then, we have the following approximation of the error:
\begin{align}
 &\epsilon(\vec{\theta}) \nonumber\\
=& \frac{1}{2} \sum_{i=1}^M \frac{\partial^2}{\partial \theta_i^2} C(\vec{\theta}) \sigma_i^2
+ 2 \sum_{\nu=1}^{M_{\mathrm{SNC}}} \frac{\partial^2}{\partial \xi_\nu^2} C(\vec{\theta}) p_\nu\nonumber\\
&\hspace{-0.4cm}+ O\left( \sum_{l=1}^{L}(E_{\max,l} - E_{0,l}) \left[\left(\sum_{i=1}^{M} \sigma_i^{2}\right)^2 + \left(\sum_{\nu=1}^{M_{\mathrm{SNC}}} p_\nu \right)^2 \right]\right),\label{e_fluc-stoch}
\end{align}
where $\xi_\nu$ is the virtual parameter associated with $\mathcal{E}_{B_\nu}$ introduced in Sec.~\ref{ssec-correspond} to give the correspondence between the stochastic noise and the Gaussian noise models.
\end{corollary}
}Especially, as a typical model, we consider a \cod{local depolarizing noise} model such that the depolarizing channel $\mathcal{D}_{k, q_k}$ is inserted after each $k$-qubit gate, where we set $q_k = (4^{k-1} - 4^{-1})c_k q$ with $q$ being the scaling of the error probability, and
$c_k$ being the constant factor characterizing the difference in the error rates between different number-qubit gates.
Let the fluctuations of the optimizing parameters themselves be negligible $\sigma_i = 0$ $(i = 1, 2, \cdots, M)$ in this case.
\cod{Under this local depolarizing noise model, the following proposition holds:
\begin{proposition}
 Let $\sum_{l=1}^{L}(E_{\max,l} - E_{0,l}) = O(n^r)$ hold with a positive number $r$.
 Under the above local depolarizing noise model, 
we can achieve a given desired precision $\epsilon_{*}$ in the sense that
\begin{align}
 \epsilon(\vec{\theta}) = O(\epsilon_{*}),
\end{align}
when the error probability has the scaling
\begin{align}
 q = O\left(\frac{\epsilon_*}{n^r M}\right).\label{pbida}
\end{align}
\end{proposition}
\begin{proof}
 Applying Corollary \ref{cor1} to the local depolarizing noise model in combination with Lemma \ref{Lem-gauss-dep}, we obtain
\begin{align}
 &\epsilon(\vec{\theta})\nonumber\\
=& \frac{1}{2} \sum_{\nu=1}^{M_{\mathrm{DP}}} \frac{\partial^2}{\partial \xi^{2}_\nu} C(\vec{\theta}) c_{k_\nu} q + O\left(\sum_{l=1}^{L}(E_{\max,l} - E_{0,l}) M^2 q^2\right),\label{err-est-q}
\end{align}
where each $\xi_\nu$ denotes the
virtual parameter
associated with each stochastic Pauli noise channel in the decomposition of one of the $k_\nu$-qubit depolarizing channels in the circuit,
and the total number of the stochastic Pauli noise channels $M_{\mathrm{DP}}$ satisfies $M_{\mathrm{DP}}=O(M)$.
Since the second derivatives are bounded as $\left|\frac{\partial^2}{\partial \xi_\nu^2} C(\vec{\theta})\right| \leq \sum_{l=1}^{L}(E_{\max,l} - E_{0,l})/2$ from Eq.~\eqref{der-bd},
the estimation
\begin{align}
 \sum_{\nu=1}^{M_{\mathrm{DP}}} \frac{\partial^2}{\partial \xi_{\nu}^{2}} C(\vec{\theta}) c_{k_\nu}
= O\left(M \sum_{l=1}^{L}(E_{\max,l} - E_{0,l})\right) = O(M n^r)\label{bioaj}
\end{align}
holds.
Hence, if $q$ satisfies Eq.~(\ref{pbida}), we obtain
\begin{align}
 \epsilon(\vec{\theta}) =& \frac{1}{2} \sum_{\nu=1}^{M_{\mathrm{DP}}} \frac{\partial^2}{\partial \xi^{2}_\nu} C(\vec{\theta}) c_{k_\nu} q + O\left(\frac{\epsilon_*^2}{n^r}\right)\nonumber\\
=& \frac{1}{2} O(M n^r) O\left(\frac{\epsilon_*}{n^r M}\right) + O\left(\frac{\epsilon_*^2}{n^r}\right)\nonumber\\
=& O(\epsilon_{*}).\label{mhaclzhc}
\end{align}
\end{proof}
} 
For example, when $r=1$, to achieve $\epsilon(\vec{\theta})\sim 10^{-3}$ (i.~e.~we set $\epsilon_* = 10^{-3}$) with $n \sim 100$ qubits and the number of gates $M \sim 100$, the error probability $q \sim 10^{-7}$ is sufficient, according to this order estimation.
\com{As we will show in Sec.~\ref{jshqlzhb},
a simple error mitigation method utilizing Theorem \ref{thm1} can relax this stringent error estimation.
We also remark that this order estimation does {\it not} mean that Eq.~(\ref{pbida}) is required to achieve the precision $\epsilon_{*}$,
but it only shows that Eq.~(\ref{pbida}) is {\it sufficient} for that.
Hence, larger error probability than this estimation might be acceptable in practice.
Another estimation to give a necessary error level will be shown in Sec.~\ref{jvfalzhb} by a lower bound (\ref{rough_LB}).
}

From another point of view, the coefficients $\frac{\partial^2}{\partial \theta^{2}_i} C(\vec{\theta})$ and $\frac{\partial^2}{\partial \xi^{2}_{\nu}} C(\vec{\theta})$ in Eq.~(\ref{e_fluc-stoch}) give the sensitivity to the noise.
\coc{Especially, low sensitivity to the fluctuations in the optimizing parameters requires small diagonal components of the Hessian of the cost function.}
For a minimal point $\vec{\theta}^*$, this means that the trace norm of the Hessian should be small for the low sensitivity to the \coc{fluctuations} since the Hessian is positive, which implies the flat landscape of the cost function around the minima.
However, the optimization in a flat landscape tends to be hard, e.~g.~due to the required precision of the gradient in gradient descent methods, which increases the required measurement number.
Hence, Eq.~(\ref{berib}) implies a trade-off relation between the sensitivity to the fluctuations and the trainability of the cost function.

\coc{The above argument can be extended to stochastic noise models when a part of the virtual parameters coincides with some optimizing parameters in the sense that $B_\nu = A_j$-stochastic noise occurs next to $U_j(\theta_j)$ gate.
For example, when all $A_j$ are Pauli operators, and depolarizing noise $\mathcal{D}_{k, q_k}$ acting on the same qubit number $k$ as $A_j$ is inserted after each $U_j(\theta_j)$, one of the stochastic Pauli noise channels composing depolarizing noise is the stochastic $A_j$-channel.
In such a case, a part of the effects of the stochastic noise can be regarded as fluctuations in optimizing parameters.
To proceed our analysis, we have to separately estimate the derivatives with respect to the virtual parameters which do not coincide with any optimizing parameters
since these virtual parameters have nothing to do with the optimization landscape.
Here, we call such virtual parameters {\it proper} virtual parameters.}
We define the noiseless precision $\delta(\vec{\theta}^*)$ of the minimization as $\delta(\vec{\theta}^*) := C(\vec{\theta}^*) - E_0$ which attributes to poor expression power of the parameterized quantum circuit $U(\vec{\theta})$ and to the non-globality of the minimization (i.~e.~$\vec{\theta}^*$ may be a local minimum).
We assume that the parameters giving the minima of the noisy cost function does not significantly deviate from the noiseless ones \cite{Sharma_2020}.

For concreteness, we consider the case where all $A_j$ are Pauli operators.
The noise model is the local depolarizing noise model, i.~e.~depolarizing noise $\mathcal{D}_{k, q_k}$ acting on the same qubit number $k$ as $A_j$ is inserted after each $U_j(\theta_j)$.
We again set $q_k = (4^{k-1} - 4^{-1})c_k q$ with the scale $q$ and the constant factor $c_k$ depending on $k$.
In this case, remind that one of the stochastic Pauli noise channels composing depolarizing noise is the stochastic $A_j$-channel.
Then, the derivative with respect to the virtual parameter associated with this channel is equivalent to the derivative with respect to the optimizing parameter $\theta_j$.
We exclude such non-proper virtual parameters, and only consider proper virtual parameters as the virtual parameters $\vec{\xi} = (\xi_1,\xi_2,\cdots,\theta_{M_{\mathrm{DP, prop}}})$, \cod{where $M_{\mathrm{DP, prop}} < M_{\mathrm{DP}}$ is the number of the proper virtual parameters.}
We note the number $M_{\mathrm{DP,prop}}$ of the proper virtual parameters is again of order $O(M)$.
Let $m_i$ be the number of qubits $A_i$ acting on.
We also define $k_\nu$ in the same way as in Eq.~(\ref{err-est-q}).
Hence, $m_i = k_{\nu}$ holds for $\nu$ with $\xi_\nu$ associated with the depolarizing channel next to the $i$-th parametric gate.
For convenience, we rescale the parameters as $\theta_i = \sqrt{c_{m_i}} \tilde{\theta}_i$.
Let $c := \max_{\nu} c_{k_\nu}$.
Then, we can prove the following proposition (see Appendix \ref{sec_prop2} for its proof):
\cod{\begin{proposition}\label{prop2}
Let all $A_j$ $(j=1,\cdots,M)$ be Pauli operators.
Under the above local depolarizing noise model, the following inequality holds
 \begin{align}
 & 2\frac{\epsilon(\vec{\theta}^*)}{q} + c M_{\mathrm{DP}}\delta(\vec{\theta}^*) + O\left(\sum_{l=1}^{L}(E_{\max,l} - E_{0,l}) M^2 q\right)
\nonumber\\
\geq &
 \Tr \left|\left(\frac{\partial^2 C}{\partial \tilde{\theta}_i \partial \tilde{\theta}_j} (\vec{\theta}^*)\right)\right|.\label{h-est}
\end{align}
\end{proposition}}
For a successful minimization, $\delta(\vec{\theta}^*)$ should be small, and hence the term $c M_{\mathrm{DP}} \delta(\vec{\theta}^*) = O(M\delta(\vec{\theta}^*))$ is negligible for \coc{not too large $M$ such that $M = o\left(\delta(\vec{\theta}^*)^{-1}\right)$.}
$O\left(\sum_{l=1}^{L}(E_{\max,l} - E_{0,l}) M^2 q\right)$ term is also negligible for sufficiently small error probability $q$.
Then, Eq.~(\ref{h-est}) implies that the trace norm of the Hessian of the cost function should be small if the error probability $q$ is not sufficiently small compared to the required level of the error $\epsilon(\vec{\theta}^*) < \epsilon_{*} $ due to the noise.
This fact implies the hardness of the optimization due to the flat landscape of the vicinity of the minima.
Moreover, in this case, optimization algorithms utilizing the Hessian become hard since high precision of the estimation of the Hessian is required if the Hessian is small.
Oppositely, at least we need $q = O(\epsilon_{*})$ to achieve $\epsilon_{*} > \epsilon(\vec{\theta}^*)$ avoiding such hardness.


\subsection{An error mitigation method}
\label{jshqlzhb}

We can apply Theorem \ref{thm1} to derive an error mitigation method.
We can cancel the error by subtracting the leading term of the error $\frac{1}{2} \sum_{i=1}^{M_{\mathrm{tot}}} \frac{\partial^2}{\partial \theta_i^2} C(\vec{\theta}) \sigma_i^2$
given that we know the error model, and $\sigma_i^2$ is small enough so that the sub-leading order terms of
$O\left(\sum_{l=1}^{L}(E_{\max,l} - E_{0,l})\left(\sum_{i=1}^{M_{\mathrm{tot}}} \sigma_i^{2}\right)^2\right)$ are negligible.
\com{An advantage of this method is that we only use the noisy estimation of the derivatives of the cost function to mitigate the error, and we do not need to change the noise strength as in the extrapolation method \cite{PhysRevX.7.021050,PhysRevLett.119.180509}, nor to sample various circuits as in the probabilistic error cancellation \cite{PhysRevLett.119.180509, PhysRevX.8.031027}.}
Using the parameter shift rule (\ref{even-d}), we can calculate the second derivatives from noisy evaluations of the cost function.
The effect of the noise in this noisy estimation $h_i(\vec{\theta})$ of the second derivative is estimated by applying Theorem \ref{thm1} again, which reads
\begin{align}
 h_i(\vec{\theta})
=& \frac{1}{2} \left[C_{\mathrm{noisy}}(\vec{\theta} + \pi \vec{e}_i) - C_{\mathrm{noisy}}(\vec{\theta})\right]\nonumber\\
=& \frac{\partial^2}{\partial \theta_i^2} C(\vec{\theta})
+ O\left(\sum_{l=1}^{L}(E_{\max,l} - E_{0,l})\sum_{i=1}^{M_{\mathrm{tot}}} \sigma_i^{2}\right).
\end{align}
\cod{Therefore, the error-mitigated cost function $C_{\mathrm{mitigated}}(\vec{\theta})$ defined as
\begin{align}
& C_{\mathrm{mitigated}}(\vec{\theta})\nonumber\\
:=& \left(1 + \frac{1}{4}\sum_{i=1}^{M_{\mathrm{tot}}}\sigma_i^2 \right)C_{\mathrm{noisy}}(\vec{\theta}) - \frac{1}{4}\sum_{i=1}^{M_{\mathrm{tot}}}C_{\mathrm{noisy}}(\vec{\theta} + \pi \vec{e}_i)\sigma_i^2
\end{align}
yields
\begin{align}
C_{\mathrm{mitigated}}(\vec{\theta})
= C(\vec{\theta}) + O\left(\sum_{l=1}^{L}(E_{\max,l} - E_{0,l})\left(\sum_{i=1}^{M_{\mathrm{tot}}} \sigma_i^{2}\right)^2\right).\label{QEM_eq}
\end{align}
Eq.~(\ref{QEM_eq}) is verified by observing that
\begin{align}
&C_{\mathrm{mitigated}}(\vec{\theta})\nonumber\\
 =&C_{\mathrm{noisy}}(\vec{\theta}) - \frac{1}{2}\sum_{i=1}^{M_{\mathrm{tot}}} h_i(\vec{\theta}) \sigma_i^2\nonumber\\
=& C(\vec{\theta}) + \frac{1}{2} \sum_{i=1}^{M_{\mathrm{tot}}} O\left(\sum_{l=1}^{L}(E_{\max,l} - E_{0,l})\sum_{i=1}^{M_{\mathrm{tot}}} \sigma_i^{2}\right) \sigma_i^2 \nonumber\\
&+ O\left(\sum_{l=1}^{L}(E_{\max,l} - E_{0,l})\left(\sum_{i=1}^{M_{\mathrm{tot}}} \sigma_i^{2}\right)^2\right)\nonumber\\
=& C(\vec{\theta}) + O\left(\sum_{l=1}^{L}(E_{\max,l} - E_{0,l})\left(\sum_{i=1}^{M_{\mathrm{tot}}} \sigma_i^{2}\right)^2\right).
\end{align}}
Hence, in this way, we can mitigate the error up to the sub-leading order $O\left(\sum_{l=1}^{L}(E_{\max,l} - E_{0,l})\left(\sum_{i=1}^{M_{\mathrm{tot}}} \sigma_i^{2}\right)^2\right)$.

This method is also applicable to the stochastic noise including depolarizing noise by applying Corollary \ref{cor1}.
The overhead of this protocol is the evaluations of the noisy cost function at the $\pi$-shift of every parameter \com{including the virtual parameters}.
\coc{$\pi$-shift of a virtual parameter $\xi_\nu$ can be implemented by actually applying its generator $B_\nu$ at the error occurs.
In the case of depolarizing noise, each Pauli rotation gate is inserted to calculate the second derivative with respect to each virtual parameter.}
\coc{Although the extra noise is added as a byproduct of this inserted gate}, the order estimation is not affected, since at most a single gate is inserted for each evaluation.
We again consider the same local depolarizing noise model with the scaling of the error probability $q$ as the one to obtain Eq.~(\ref{err-est-q}).
We also assume that $\sum_{l=1}^{L}(E_{\max,l} - E_{0,l}) = O(n^r)$.
Then, in order to achieve a given precision $\epsilon_*$, it is sufficient to have
\begin{align}
 q = O\left(\frac{\epsilon_*^{\frac{1}{2}}}{n^{\frac{r}{2}} M}\right)
\end{align}
by applying this error mitigation.
In comparison to Eq.~(\ref{pbida}), the order estimation of the sufficient noise level is relaxed by $\sqrt{\epsilon_* / n^r}$ via this error mitigation.
For example, when $r = 1$, to achieve the precision $\sim 10^{-3}$ with $n\sim 100$ qubits and the number of gates $M\sim 100$, the error probability $q\sim 3 \times 10^{-5}$ is sufficient, which is about $10^2$ times larger in comparison with the one without the error mitigation shown below Eq.~\eqref{mhaclzhc}, although it is still stringent.
However, we again remark that this estimation is only the sufficient order of the error probability to achieve a given precision, but not necessary.
Moreover, we can take into account the next-leading order in expansion (\ref{ithelzgp}) to improve the error mitigation if the overhead is acceptable.
Further analysis on the practical effectiveness of this error mitigation method including the finiteness of the sampling and the comparison with different error mitigation techniques will be done in a successive work.

\section{Lower and upper bounds of the precision}\label{jvfalzhb}
\coc{In this section, we focus on the deviation $\epsilon_0(\vec{\theta}):= C_{\mathrm{noisy}}(\vec{\theta}) - E_0$ of the noisy cost function $C_{\mathrm{noisy}}(\vec{\theta})$ from the minimum eigenvalue $E_0$ as the error of the noisy VQA task to estimate $E_0$.
We show upper and lower bounds of $\epsilon_0(\vec{\theta})$.
}
The bounds reveal how the spectrum of the target operator and the geometric structure of the ansatz affect the sensitivity of the cost function to the noise.
Especially, from the lower bound, we can estimate how small error probability is required to achieve a given precision under reasonable assumptions.
We can also derive rough estimations of the bounds which can be easy to check,
instead of calculating the Hessian of the cost function in (\ref{eq-thm1}), which would be too expensive to calculate just for the error estimations.

In the following, we focus on the case where all the input states $\ket{\phi_l}$ $(l = 1,2,\cdots,L)$ are the same as $\ket{\phi_l} = \ket{\phi}$.
In this case the cost function is reduced to the expectation value of a single Hermitian operator $H = \sum_{l=1}^{L} H_l$ since $C(\vec{\theta}) = \sum_{l=1}^{L}\bra{\phi} U(\vec{\theta})^{\dagger} H_l U(\vec{\theta}) \ket*{\phi} = \bra{\phi} U(\vec{\theta})^{\dagger} H U(\vec{\theta}) \ket*{\phi}$.
We denote the smallest, the second smallest, and the maximum eigenvalues of $H$ by $E_0$, $E_1$, and $E_{\max}$, respectively.
We assume that the eigenspace for the minimum eigenvalue $E_0$ of $H$ is nondegenerate.

In the following analysis, we proceed based on the stochastic noise model. Especially, Gaussian fluctuations in the optimizing parameters can also be modeled as the stochastic noise $\mathcal{E}_{A_i, p_i}$ with respect to the generator $A_i$ of $U_i(\theta_i)$ through the correspondence shown in Sec.~\ref{ssec-correspond}, where the error probability $p_i$ is given as $p_i = [1 - \exp(-\sigma_i^2/2)]/2$.
Moreover, the action of $A_i$-error in the circuit to the cost function is the same as the shift of $\theta_i$ by $\pi$.
$B_{\nu}$-error of the stochastic noise can also be represented as the shift of the virtual parameter $\xi_{\nu}$ by $\pi$.
Hence, for convenience, we treat the optimizing and virtual parameters together in the same notation in the same way as in Sec.~\ref{jvxylzhd}.
Now, we introduce the quantity $G_{i_1, i_2, \cdots, i_k} (\vec{\theta})$ which describes the sensitivity of the state $\ket*{\phi(\vec{\theta})} := U(\vec{\theta})\ket{\phi}$ to the $\pi$-shift of the parameters $\theta_{i_1}, \cdots, \theta_{i_k}$ as follows:
\begin{align}
 G_{i_1, i_2, \cdots, i_k} (\vec{\theta}) := 1 - \left|\bra{\phi\left(\vec{\theta} + \pi\sum_{l=1}^k\vec{e}_{i_l}\right)}\ket{\phi(\vec{\theta})}\right|^2\label{Giii}
\end{align}
with $k = 1,2,\cdots,M_{\mathrm{tot}}$.
In particular, $G_i(\vec{\theta})$ $(i=1,\dots, M)$ corresponds to the diagonal components of the Fubini-Study metric of the ansatz states
\begin{align*}
 g_{i,j}(\vec{\theta}) := &
\mathrm{Re}\Bigg[\braket{\frac{\partial}{\partial \theta_i} \phi (\vec{\theta})}{\frac{\partial}{\partial \theta_j} \phi (\vec{\theta})} \nonumber\\
& - \braket{\frac{\partial}{\partial \theta_i} \phi (\vec{\theta})}{\phi(\vec{\theta})}\braket{\phi(\vec{\theta})}{\frac{\partial}{\partial \theta_j} \phi (\vec{\theta})} \Bigg]
\end{align*}
with the relation $G_i(\vec{\theta}) = 4g_{i,i}(\vec{\theta})$.
Then, we obtain the following lower and upper bounds of the error, \cod{which are proved in Appendix~\ref{sec_thm_bounds}}:
\begin{theorem}\label{thm2}
Let all the error probabilities satisfy $p_i < 1$ $(i=1,\cdots, M_{\mathrm{tot}})$.
Then, the error $\epsilon_0(\vec{\theta})$ is lower bounded as
\begin{align}
 &\epsilon_0(\vec{\theta})\nonumber\\
\geq &
(E_1 - E_0) \prod_{j = 1}^{M_{\mathrm{tot}}} (1 - p_j) \nonumber\\
&\times \sum_{k=1}^{M_{\mathrm{tot}}}\sum_{i_1 \neq i_2 \neq \cdots \neq i_k} 
\frac{\prod_{l=1}^k p_{i_l}}{\prod_{l=1}^k (1-p_{i_l})}
G_{i_1, i_2, \cdots, i_k} (\vec{\theta}) + R_{\mathrm{L}}(\vec{\theta}),\label{low_bd1}
\end{align}
where
\begin{align}
 &R_{\mathrm{L}}(\vec{\theta}) \nonumber\\
:=& \prod_{i=1}^{M_{\mathrm{tot}}}(1-p_i)\delta(\vec{\theta}) - 2 \left[1 - \prod_{i=1}^{M_{\mathrm{tot}}}(1-p_i)\right] \sqrt{(E_1 - E_0) \delta(\vec{\theta})}\label{RL}
\end{align}
and $\delta(\vec{\theta}) = C(\vec{\theta}) - E_0$ is the noiseless precision.
Similarly, $\epsilon_0(\vec{\theta})$ is upper bounded as
\begin{align}
 &\epsilon_0(\vec{\theta})\nonumber\\
\leq &
(E_{\max} - E_0) \prod_{j = 1}^{M_{\mathrm{tot}}} (1 - p_j) \nonumber\\
&\times \sum_{k=1}^{M_{\mathrm{tot}}}\sum_{i_1 \neq i_2 \neq \cdots \neq i_k} 
\frac{\prod_{l=1}^k p_{i_l}}{\prod_{l=1}^k (1-p_{i_l})}
G_{i_1, i_2, \cdots, i_k} (\vec{\theta}) + R_{\mathrm{U}}(\vec{\theta}),\label{up_bd1}
\end{align}
where
\begin{align}
 R_{\mathrm{U}}(\vec{\theta})
:=& \prod_{i=1}^{M_{\mathrm{tot}}}(1-p_i)\delta(\vec{\theta})\nonumber\\
&+ 2\left[1 - \prod_{i=1}^{M_{\mathrm{tot}}}(1-p_i)\right]\frac{E_{\max} - E_0}{\sqrt{E_1 - E_0}} \sqrt{\delta(\vec{\theta})}.\label{RU}
\end{align}
\end{theorem}
Especially, let us apply Theorem \ref{thm2} at a minimal point $\vec{\theta}^*$.
Here, we again assume that the parameters giving the minima of the noisy cost function does not significantly deviate from the noiseless ones \cite{Sharma_2020}.
Then, the noiseless precision $\delta(\vec{\theta}^{*})$ should be small enough for a successful minimization, and hence
the terms $R_{\mathrm{L}}(\vec{\theta}^*)$ and $R_{\mathrm{U}}(\vec{\theta}^*)$ are negligible.
\coc{Then, the bounds (\ref{low_bd1}) and (\ref{up_bd1}) are characterized by the spectrum of $H$ and the sensitivity of the ansatz $G_{i_1,i_2,\cdots,i_k}(\vec{\theta}^*)$ as
\begin{align}
 &
(E_1 - E_0) \prod_{j = 1}^{M_{\mathrm{tot}}} (1 - p_j) \nonumber\\
&\times \sum_{k=1}^{M_{\mathrm{tot}}}\sum_{i_1 \neq i_2 \neq \cdots \neq i_k} 
\frac{\prod_{l=1}^k p_{i_l}}{\prod_{l=1}^k (1-p_{i_l})}
G_{i_1, i_2, \cdots, i_k} (\vec{\theta}^*) \nonumber\\
\lesssim& \epsilon_0(\vec{\theta}^*)\nonumber\\
\lesssim&
(E_{\max} - E_0) \prod_{j = 1}^{M_{\mathrm{tot}}} (1 - p_j) \nonumber\\
&\times \sum_{k=1}^{M_{\mathrm{tot}}}\sum_{i_1 \neq i_2 \neq \cdots \neq i_k} 
\frac{\prod_{l=1}^k p_{i_l}}{\prod_{l=1}^k (1-p_{i_l})}
G_{i_1, i_2, \cdots, i_k} (\vec{\theta}^*).\label{main_ULbounds}
\end{align}}
It is noted that $E_{\max}$ in the upper bound can be replaced with the maximum eigenvalue of the eigenspace accessible by the ansatz, according to the derivation of the bound.
Hence, the upper bound implies how the strategy that restricting the expressiveness of the ansatz can be beneficial for reducing the error due to the noise.
From the lower bound, larger gap $E_1 - E_0$ implies the larger error.
\com{As it is considered that larger spectral gap is a key to relax the computational complexity of estimating the ground state energy in general \cite{PRXQuantum.3.040327}, which is actually the case in some cases \cite{Hastings_2007,Landau:2015tb,Arad:2017ts},}
this fact implies a trade-off between the hardness of the optimization and the sensitivity to the noise.
%
%

\coc{Although it is impractical to calculate all $2^{M_{\mathrm{tot}}}$ terms with $G_{i_1, i_2, \cdots, i_k}(\vec{\theta}^*)$ $(k=1,2,\cdots,M_{\mathrm{tot}})$ in Eq.~(\ref{main_ULbounds}),
we obtain the following rough bounds up to the terms with $k=1$ because $0\leq G_{i_1,i_2,\cdots,i_k}(\vec{\theta}) \leq 1$ holds for any $\vec{\theta}$:
\begin{align}
 &
(E_1 - E_0) \sum_{i=1}^{M_{\mathrm{tot}}} p_i \prod_{j\neq i} (1-p_{j})
G_{i} (\vec{\theta}^*) \nonumber\\ 
\lesssim& \epsilon_0(\vec{\theta}^*)\nonumber\\
\lesssim&
(E_{\max} - E_0)\sum_{i=1}^{M_{\mathrm{tot}}} p_i \prod_{j\neq i} (1-p_{j})
G_{i} (\vec{\theta}^*)\nonumber\\
&+(E_{\max} - E_0)
\left[1 - \prod_{j=1}^{M_{\mathrm{tot}}}(1 - p_j)
-\sum_{i=1}^{M_{\mathrm{tot}}} p_i \prod_{j\neq i} (1-p_{j})\right],\label{R1_BDs} 
\end{align}
where we have used the fact that
\begin{align}
 &\prod_{j=1}^{M_{\mathrm{tot}}}(1 - p_j) + \prod_{j = 1}^{M_{\mathrm{tot}}} (1 - p_j) \sum_{k=1}^{M_{\mathrm{tot}}}\sum_{i_1 \neq i_2 \neq \cdots \neq i_k} 
\frac{\prod_{l=1}^k p_{i_l}}{\prod_{l=1}^k (1-p_{i_l})}\nonumber\\
=&
\prod_{j=1}^{M_{\mathrm{tot}}}\left[(1-p_j) + p_j\right]
= 1
\end{align}
to derive the second inequality.
\cod{We remark that $E_{\max} - E_0 \leq 2\|H\|$ holds for the operator norm $\|H\|$ of $H$.
Hence, Eq.~(\ref{R1_BDs}) is still available by replacing $E_{\max} - E_0$ with an upper bound of $2\|H\|$ even if we have no access to the exact value of $E_{\max} - E_0$. For example, if $H$ is decomposed into a linear combination of Pauli operators as $H = \sum_i c_i P_i$, we can use a bound $\|H\|\leq \sum_{i}|c_i|$.}
The calculation of $G_i(\vec{\theta})$ is not so expensive.
In fact, if we assume that the (virtual and optimizing) parameters $\theta_i$ are sorted in ascending order of application of their corresponding parametric gates, then $|\bra*{\phi(\vec{\theta} + \pi\vec{e}_{i})}\ket*{\phi(\vec{\theta})}|^2$ can be calculated by the shallowed circuit up to $i$-th gate because the gates after it are canceled in the inner product.
Collecting the first-order terms with respect to the error probabilities $p_i$ in Eq.~(\ref{main_ULbounds}), we obtain the respective leading-order terms $(E_1 - E_0)\sum_{i=1}^{M_{\mathrm{tot}}}p_i G_i(\vec{\theta}^*)$ and $(E_{\max} - E_0)\sum_{i=1}^{M_{\mathrm{tot}}}p_i G_i(\vec{\theta}^*)$ of the lower and the upper bounds in Eq.~(\ref{main_ULbounds}).
We remark that rough bounds (\ref{R1_BDs}) include these leading-order terms and only drop a part of the higher-order terms from Eq.~(\ref{main_ULbounds}).
Hence, rough bounds (\ref{R1_BDs}) captures the main part of Eq.~(\ref{main_ULbounds}) for small error probabilities.
As the coefficients $G_i(\vec{\theta}^*)$ of the leading-order terms coincide with the Fubini-Study metric,
the geometric structure of the ansatz is connected to the sensitivity of the cost function to the noise.
Especially, the bounds implies the following trade-off relation.}
Although small $G_{i}(\vec{\theta}^*)$ is better for the noise sensitivity, it becomes hard to calculate the metric itself, which implies the hardness of the metric aware optimization methods such as the natural gradient \cite{10.1162/089976698300017746,Stokes2020quantumnatural,1909.05074}.
On the other hand, it was shown that the average convergence speed in terms of the optimization steps of SGD can be faster by the smaller metric \cite{gentini_noise-resilient_2020}.
This result implies a possibility that the small metric simultaneously improves both the sensitivity to the noise and the convergence speed of the optimization.
However, the flat landscape due to the small metric may have rather a bad effect for the optimization due to the high precision required to determine the gradient.
In fact, the measurement number and the variance are not taken into account in the analysis of the convergence speed in Ref.~\cite{gentini_noise-resilient_2020}.

\cod{We also have the following rougher upper bound:
\begin{align}
 \epsilon_0(\vec{\theta}^*)
\lesssim
(E_{\max} - E_0) \left[1 - \prod_{j=1}^{M_{\mathrm{tot}}}(1 - p_j)\right],
\label{b8n}
\end{align}
which is derived by further applying $G_{i}(\vec{\theta}) \leq 1$ to the rightmost side Eq.~(\ref{R1_BDs}) and observing that
\begin{align}
 &\epsilon_0(\vec{\theta}^*)\nonumber\\
\lesssim &
(E_{\max} - E_0) \prod_{j = 1}^{M_{\mathrm{tot}}} (1 - p_j) \sum_{k=1}^{M_{\mathrm{tot}}}\sum_{i_1 \neq i_2 \neq \cdots \neq i_k} 
\frac{\prod_{l=1}^k p_{i_l}}{\prod_{l=1}^k (1-p_{i_l})}\nonumber\\
=&
(E_{\max} - E_0) \left[1 - \prod_{j=1}^{M_{\mathrm{tot}}}(1 - p_j)\right].
\end{align}
Eq.~(\ref{b8n}) is also available by replacing $E_{\max} - E_0$ with an upper bound of $2\|H\|$ if accessible.
In this way, we can use Eq.~(\ref{b8n}) for an easy check of the impact of the noise under a given error probabilities $p_i$ by using only accessible quantities.}

\coc{As an another approach to roughly estimating the lower bound,
it is reasonable to assume that
most of $G_{i_1,i_2,\cdots,i_k}(\vec{\theta})$ are not close to zero
since the state should considerably change as $k$ gates are inserted in the circuit, unless some specific structure exists.
Then, assuming that $G_{i_1,i_2,\cdots,i_k}(\vec{\theta}^*) > c$ holds for some constant $c > 0$, we have
\begin{align}
  &\epsilon_0(\vec{\theta}^*)\nonumber\\
\gtrsim &
(E_1 - E_0) \prod_{j = 1}^{M_{\mathrm{tot}}} (1 - p_j) \sum_{k=1}^{M_{\mathrm{tot}}}\sum_{i_1 \neq i_2 \neq \cdots \neq i_k} 
\frac{\prod_{l=1}^k p_{i_l}}{\prod_{l=1}^k (1-p_{i_l})} c\nonumber\\
=&
(E_1 - E_0) \left[1 - \prod_{j=1}^{M_{\mathrm{tot}}}(1 - p_j)\right]c.\label{rough_LBc}
\end{align}
Especially, if we assume that $c\approx 1$, we can roughly estimate a lower bound of the precision as
\begin{align}
  \epsilon_0(\vec{\theta}^*)
\gtrsim (E_1 - E_0) \left[1 - \prod_{j=1}^{M_{\mathrm{tot}}}(1 - p_j)\right].\label{rough_LB}
\end{align}
If we know some estimation or lower bound of $E_1 - E_0$, we can use Eq.~(\ref{rough_LB}) for an easy estimation of a lower bound of the precision.
It should be noted that Eq.~(\ref{rough_LB}) is not always true even if $\delta(\vec{\theta}^*)$ is small since it is based on the assumption $G_{i_1,i_2,\cdots,i_k}(\vec{\theta}^*)\approx 1$, unlike upper bound (\ref{b8n}).

From the roughly estimated lower bound (\ref{rough_LBc}), in order to achieve a given precision $\epsilon_{*}$, the error probabilities need to satisfy
\begin{align}
 \prod_{j=1}^{M_{\mathrm{tot}}}(1 - p_j)
\geq
1 - \frac{\epsilon_{*}}{c(E_1 - E_0)}.\label{cond_lower}
\end{align}
In practice, the error in a part of the gates dominates that of the others, e.~g.~the error in the two-qubit gates usually dominates the single-qubit error.
\cod{In such a case, the following proposition holds:
\begin{proposition}
 Let the error probability $p_j$ of $M_{\mathrm{dom}}$ stochastic noise channels out of $M_{\mathrm{tot}}$ dominate the others, and scale with $p$ as $p_j= \Theta(p)$.
Then, in order to achieve a given precision $\epsilon_{*}$ with $\epsilon_{*} < E_1 - E_0$, the scale $p$ of the error probability must satisfy
\begin{align}
  p \lesssim & 1 - \left(1 - \frac{\epsilon_{*}}{c(E_1 - E_0)}\right)^{\frac{1}{M_{\mathrm{dom}}}}\nonumber\\
=& \frac{\epsilon_{*}}{c(E_1 - E_0) M_{\mathrm{dom}}} + O\left(\frac{\epsilon_{*}^2}{(E_1 - E_0)^2 M_{\mathrm{dom}}}\right).\label{nec_order1}
\end{align}
\end{proposition}
\begin{proof}
Under the assumptions, condition~(\ref{cond_lower}) to achieve the precision $\epsilon_{*}$ reads
 \begin{align}
 (1-p)^{M_{\mathrm{dom}}} \gtrsim 1 - \frac{\epsilon_{*}}{c(E_1 - E_0)}.
\end{align}
Hence, for $\epsilon_{*} < E_1 - E_0$, the scale of the error rate $p$ must satisfy Eq.~(\ref{nec_order1}).
\end{proof}
Eq.~(\ref{nec_order1}) gives an order estimation of a necessary error level to achieve a desired precision $\epsilon_{*}$.
Especially, if $E_1 - E_0 = \Omega(1)$, in order to achieve the precision $\epsilon_{*}$, we need
\begin{align}
 p = O\left(\frac{\epsilon_{*}}{M_{\mathrm{dom}}}\right).
\end{align}}
This analysis can be straightforwardly applied to the local depolarizing noise model with the scaling of the error probability $q$ as the one to obtain Eq.~(\ref{err-est-q}).
For example,
the error rate $q\sim 10^{-5}$ or less is required when $M_{\mathrm{dom}}\sim 100$, and $\epsilon_* \sim 10^{-3}$}, which is $10^2$ times larger than the sufficient order to achieve the same precision shown below Eq.~(\ref{mhaclzhc}).
This stringent requirement seems reasonable without any error mitigation.

\section{Numerical simulation}\label{sec-numerics}

\begin{figure*}
\centering
 \includegraphics[clip ,width=7.0in]{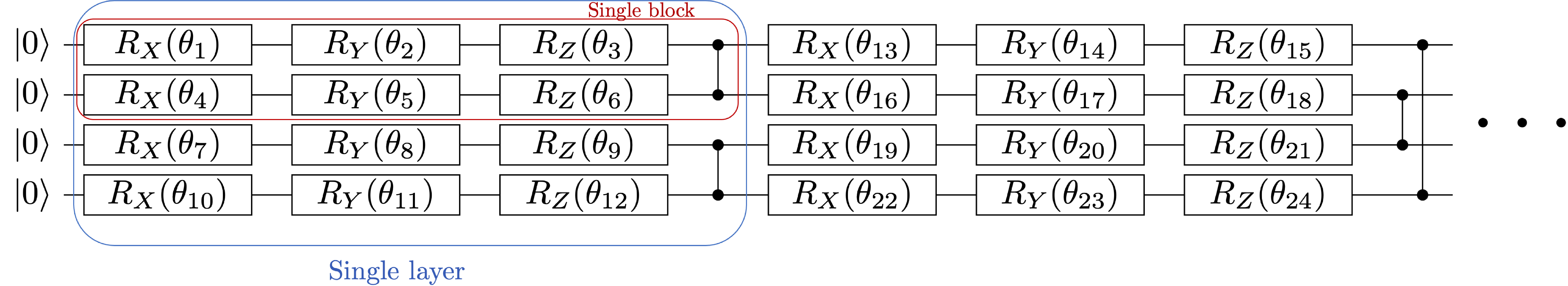}
 \caption{\coc{The $4$-qubit alternating layered circuit used in our numerical simulations. Each layer of the circuit has separated entangling blocks. A single layer is composed of single-qubit Pauli $X, Y, Z$ rotations followed by entanglers composed of the controlled-$Z$ gates acting on the adjacent qubits within a single entangling block. Each entangling block moves to the next pair of qubits when moving to the next layer, where the boundary qubits are connected.}}
\label{fig-circuit}
\end{figure*}
\begin{figure}
\centering
 \includegraphics[clip ,width=3.4in]{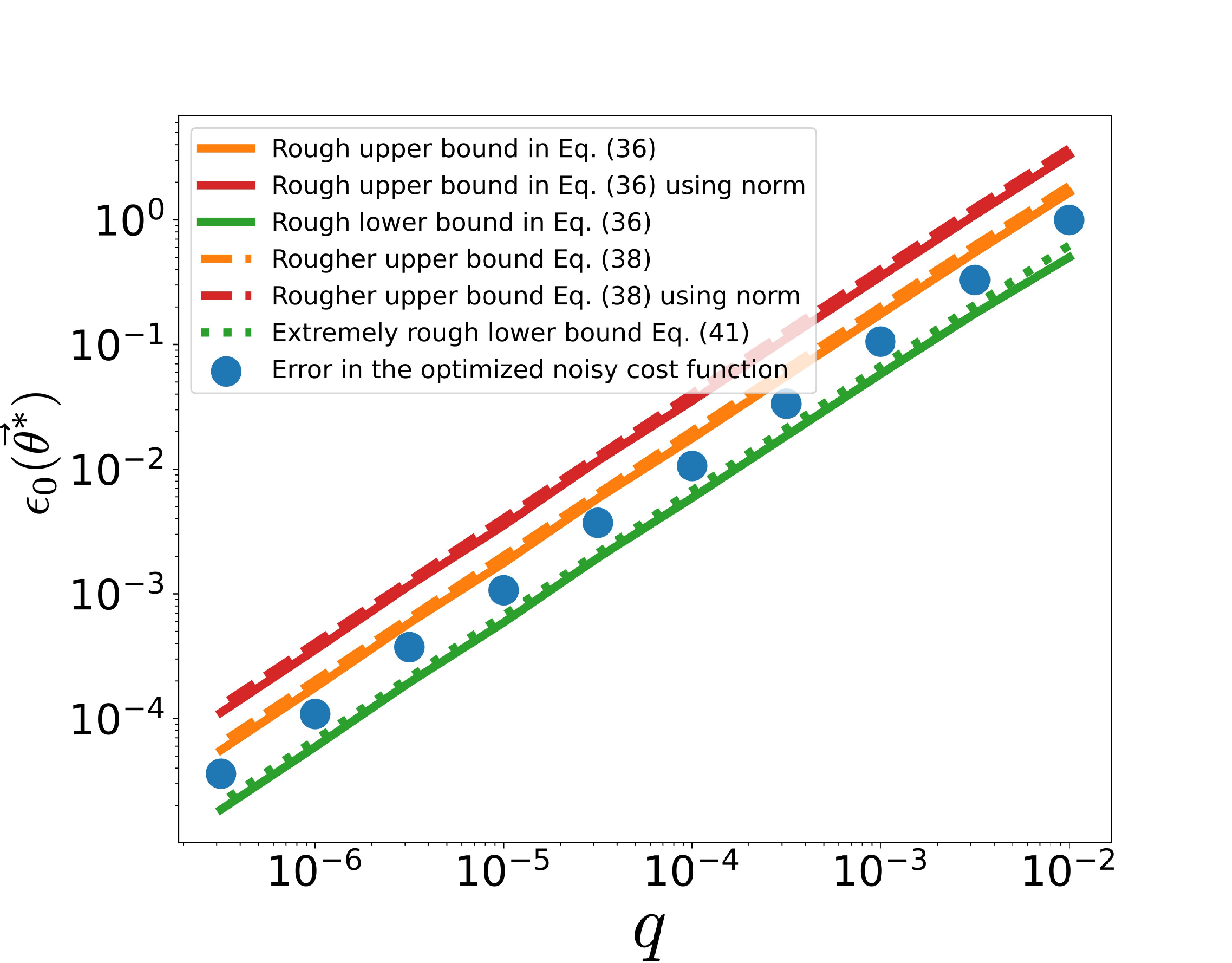}
 \caption{Dependence of the error $\epsilon_0(\vec{\theta}^*)$ in the optimized noisy cost function on the error rate $q$ for the Heisenberg spin chain. Here, the two-qubit error probability and the readout error probability are $q$, and the single-qubit error probability is $10^{-1}q$. The rough bounds are compared with the obtained values of the error $\epsilon_0(\vec{\theta}^*)$ indicated by the blue dots. The solid lines indicate the rough bounds in Eq.~(\ref{R1_BDs}). The dashed line indicates rougher upper bound Eq.~(\ref{b8n}). The dotted line indicates extremely rough lower bound Eq.~(\ref{rough_LB}). \cod{Upper bounds (\ref{R1_BDs}) and (\ref{b8n}) calculated by using bound (\ref{norm_bd}) are shown by red lines.}}
\label{fig-rate_HAF}
\end{figure}
\begin{figure}
\centering
 \includegraphics[clip ,width=3.4in]{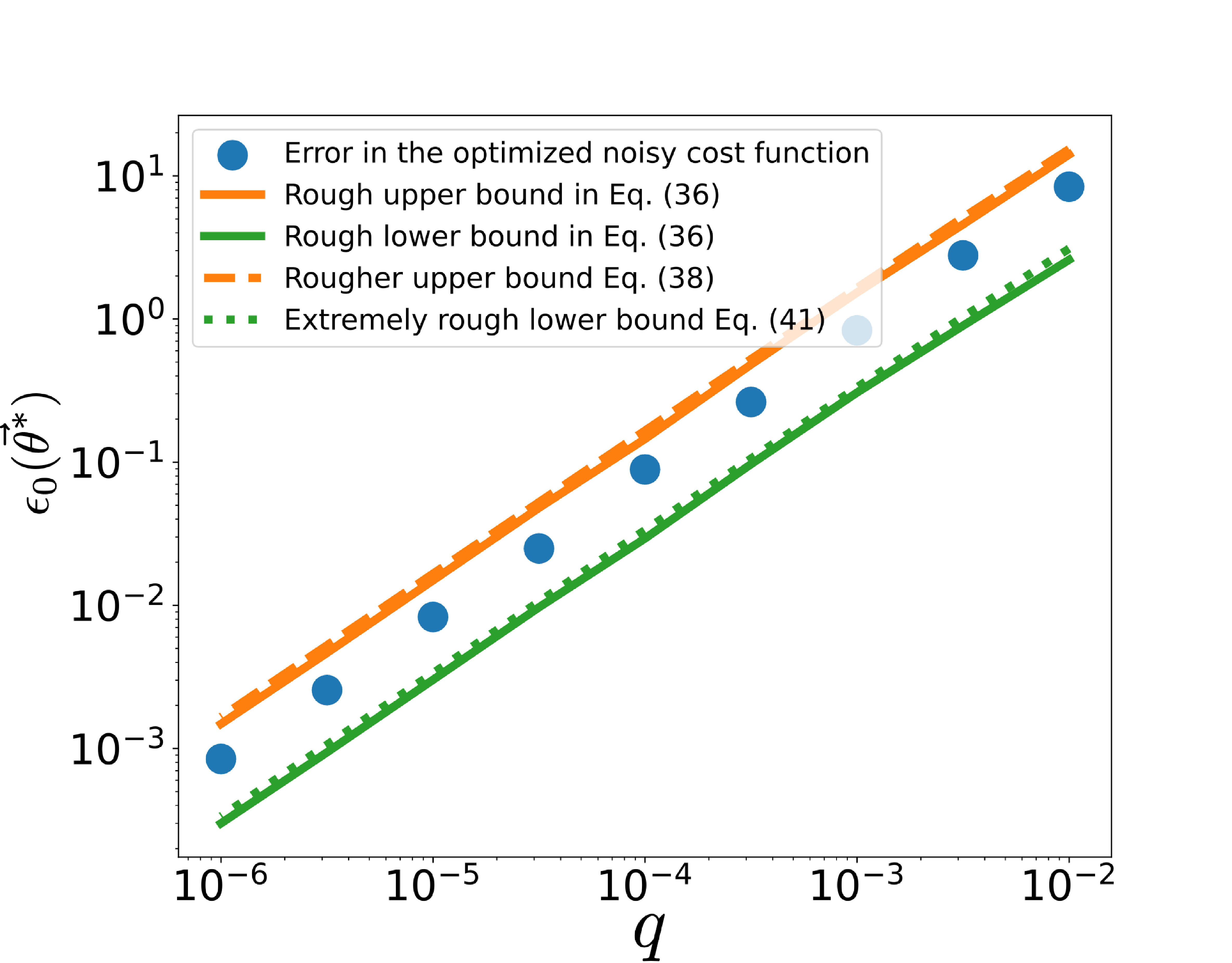}
 \caption{Dependence of the error $\epsilon_0(\vec{\theta}^*)$ in the optimized noisy cost function on the error rate $q$. Here, the two-qubit error probability and the readout error probability are $q$, and the single-qubit error probability is $10^{-1}q$. $E_1 - E_0$ is set to $50$. The rough bounds are compared with the obtained values of the error $\epsilon_0(\vec{\theta}^*)$ indicated by the blue dots. The solid lines indicate the rough bounds in Eq.~(\ref{R1_BDs}). The dashed line indicates rougher upper bound Eq.~(\ref{b8n}). The dotted line indicates extremely rough lower bound Eq.~(\ref{rough_LB}).}
\label{fig-rate}
\end{figure}
\begin{figure}
\centering
 \includegraphics[clip ,width=3.5in]{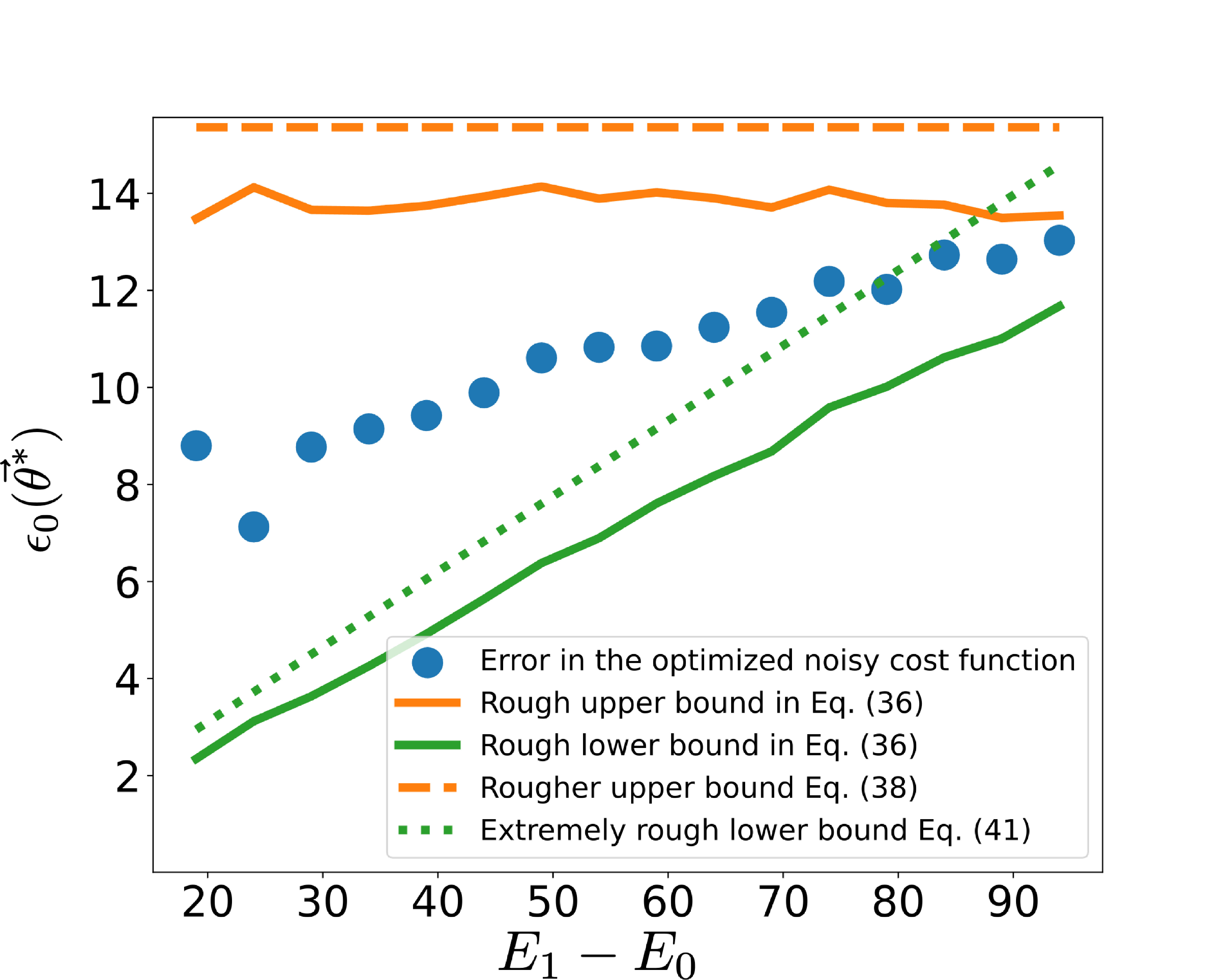}
 \caption{Dependence of the error $\epsilon_0(\vec{\theta}^*)$ in the optimized noisy cost function on the gap $E_1 - E_0$ \com{with $E_0=1.0$, $E_{\max} = 100$.} The error probability is set to $10^{-3}$ for the single-qubit error, and $10^{-2}$ for the two-qubit error and the readout error. The rough bounds are compared with the obtained values of the error $\epsilon_0(\vec{\theta}^*)$ indicated by the blue dots. The solid lines indicate the rough bounds in Eq.~(\ref{R1_BDs}). The dashed line indicates rougher upper bound Eq.~(\ref{b8n}). The dotted line indicates extremely rough lower bound Eq.~(\ref{rough_LB}).}
\label{fig-gap}
\end{figure}

We demonstrate our results by numerical simulations using Qulacs \cite{Suzuki2021qulacsfast}.
\coc{Especially, we focus on rough bounds (\ref{R1_BDs}), (\ref{b8n}) and (\ref{rough_LB}) which can be practically accessible.
We remark that all of these bounds are not exact because they are obtained by neglecting the terms $R_{\mathrm{L}}(\vec{\theta})$ and $R_{\mathrm{U}}(\vec{\theta})$ in the exact bounds in Theorem \ref{thm2}.
That approximation is based on assuming that the noiseless precision $\delta(\vec{\theta})$ is small enough.
To reflect this fact, we call the upper and the lower bounds in Eq.~(\ref{R1_BDs}) ``rough'' upper and lower bounds respectively.
We call the upper bound (\ref{b8n}) ``rougher'' upper bound, as it is rougher than the rough upper bound in Eq.~(\ref{R1_BDs}).
On the other hand, we call lower estimation (\ref{rough_LB}) ``extremely rough'' lower bound, as it can be violated even if the noiseless precision is small enough because it is based on the additional rough assumption $G_{i_1,i_2,\cdots,i_k}(\vec{\theta}^*)\approx 1$.
To distinguish among these bounds with different degree of roughness, we indicate the bounds in Eq.~(\ref{R1_BDs}) by solid lines and rougher upper bound (\ref{b8n}) by dashed lines and extremely rough lower bound (\ref{rough_LB}) by dotted lines in Figs.~\ref{fig-rate_HAF}, \ref{fig-rate}, and \ref{fig-gap}.

We implement our simulations for VQE tasks of $4$-qubit Heisenberg antiferromagnetic spin chain and a toy model Hamiltonian associated with a variational compiling task \cite{nakanishi_sequential_2020}.
For both simulations, we use an alternating layered ansatz (ALT) \cite{Cerezo:2021vi} shown in Fig.~\ref{fig-circuit} as our parameterized quantum circuit $U(\vec{\theta})$, where $R_P(\theta_i)= \exp[- i \theta_{i} P/2]$ is a single-qubit Pauli rotation gate with $P=X, Y, Z$. As explained in detail in the caption of Fig.~\ref{fig-circuit}, ALT has some entangling blocks in each layer which are alternated layer by layer.
It has been shown that ALT has both good expressibility and trainability \cite{Nakaji2021expressibilityof}, which motivates our choice of the ansatz.
For both models, we implement our simulations for the circuit with $4$ layers.
As for the model of the noise, the single(two)-qubit depolarizing channel is inserted after every single(two)-qubit gate.
We call these errors single(two)-qubit errors.
The single-qubit depolarizing channel on every qubit is also inserted after the final layer as a model of the imperfection of the measurements which we call the readout error.
In the following, the single(two)-qubit or readout error probability $p$ refers to the error probability $p$ of the corresponding depolarizing channel $\mathcal{D}_{k, p}$ defined in Eq.~(\ref{DP_def}).
\cod{To calculate bounds (\ref{R1_BDs}), (\ref{b8n}) and (\ref{rough_LB}), we decompose $k$-qubit depolarizing channels into $4^k - 1$ stochastic channels and use $\tilde{p}$ given by Eq.~(\ref{eq-pauli-d}).}
For simplicity, we do not treat fluctuations in the optimizing parameters.
In our simulations, we exactly calculated the noisy cost function by calculating the density matrix of the noisy circuits.
To obtain a minimal point of the noisy cost function, we first find a good minimizer of the noiseless cost function by the Broyden–Fletcher–Goldfarb–Shanno (BFGS) algorithm \cite{broyden1970convergence,10.1093/comjnl/13.3.317,goldfarb1970family,shanno1970conditioning} \com{via SciPy \cite{Jones:2001uv}} starting from a randomly chosen initial parameters.
We repeat the above optimization until a good solution is reached to avoid becoming stuck in local minima.
Minimization of the noisy cost functions is then done by the BFGS algorithm using this good parameter as the initial parameter.
Although the above approach of course does not work in practice,
we used it because our purpose is to demonstrate our bounds to estimate the precision of the noisy cost function to approximate $E_0$ at an in-principle-achievable good parameter.}
\subsection{Heisenberg spin chain}
We consider the VQE of $4$-qubit Heisenberg antiferromagnetic spin-$1/2$ chain with periodic boundary condition whose Hamiltonian is given as
\begin{align}
 H = \sum_{i=1}^4 (X_i X_{i+1} + Y_{i} Y_{i+1} + Z_{i} Z_{i+1}),
\end{align}
where $P_i$ $(P=X,Y,Z)$ is the Pauli operator acting on $i$-th qubit with the identification $P_{5} = P_1$.
Our task is to obtain a good parameter $\vec{\theta}^*$ to approximate the ground state energy $E_0$ of this $H$ by $C_{\mathrm{noisy}}(\vec{\theta}^*)$.
We used an ALT in Fig.~\ref{fig-circuit} with 4 layers as our ansatz.
The ALT we used can actually achieve very good noiseless precision $\delta(\vec{\theta}^*) < 10^{-6}$ which is verified by optimizing the noiseless cost function using BFGS algorithm.
Fig.~\ref{fig-rate_HAF} shows the dependence of the error $\epsilon_0(\vec{\theta}^*)$ on the error rate $q$ in comparison with rough bounds (\ref{R1_BDs}), (\ref{b8n}) and (\ref{rough_LB}).
Here, the error probability of the two-qubit error and the readout error are $q$, and the error probability of the single-qubit error is $10^{-1}q$.
To calculate the bounds, we used the exact values $E_1 - E_0 = 4.000$ and $E_{\max} - E_0 = 12.000$ for the $4$-qubit Heisenberg spin chain.
\cod{More practically, we can use the bound
\begin{align}
 &E_{\max} - E_0 \nonumber\\
\leq& 2\|H\|
\leq 2\sum_{i=1}^{4}(\|X_i X_{i+1}\| + \|Y_{i} Y_{i+1}\| + \|Z_{i} Z_{i+1}\|) \nonumber\\
& \hspace{1.2cm}= 24\label{norm_bd}
\end{align}
 to compute upper bounds (\ref{R1_BDs}), (\ref{b8n}).
We also show the bounds using this upper bound of $\|H\|$ by red lines in Fig.~\ref{fig-rate_HAF}.}
Our rough bounds actually well capture the scaling of the true error dependence on the error rate.
Especially, extremely rough lower bound~(\ref{rough_LB}) works well as a lower bound of $\epsilon_0(\vec{\theta}^*)$,
despite the fact that Eq.~(\ref{rough_LB}) is not always true as it is based on the rough assumption $G_{i_1,i_2,\cdots,i_k}(\vec{\theta}^*)\approx 1$ which can be violated.
This behavior is expected from the fact that the deviation of Eq.~(\ref{rough_LB}) from true lower bound~(\ref{main_ULbounds}) caused by the deviation of $G_{i_1,i_2,\cdots,i_k}(\vec{\theta}^*)$ from $1$ is suppressed if the error probabilities $p_i$ are small and the gap $E_1 - E_0$ is not so large.
We also remark that bounds~(\ref{R1_BDs}) which are always true under the small $\delta(\vec{\theta}^*)$ similarly work well.

\subsection{Toy model}
Next, in a similar way to \cite{nakanishi_sequential_2020}, we consider a toy target operator which has an exact solution parameter $\vec{\theta}^*$ which is uniformly and randomly selected.
That is, we generate artificial eigenstates by $\ket{\psi_i} := U(\vec{\theta}^*)\ket{i}$ $(i=0,1,\cdots, 2^n - 1)$ for a given ansatz circuit $U(\vec{\theta})$, where $\ket{i} = \ket{i_1}\otimes \ket{i_2} \otimes \cdots \otimes \ket{i_n}$ with the binary expansion $i = i_1 i_2 \cdots i_n$ for $i$.
We fix the smallest, the next smallest and the maximum eigenvalues $E_0$, $E_1$ and $E_{\max} = E_{2^n - 1}$ respectively.
The remaining eigenvalues between $E_1$ and $E_{2^n - 1}$ are randomly selected.
Then, the target operator is
\begin{align}
 H = \sum_{i = 0}^{2^n - 1} E_i \ketbra{\psi_i}.
\end{align}
In this way, the noiseless precision $\delta(\vec{\theta}^*) = 0$ can be always satisfied.
In our simulation, we consider the case where $E_0 = 1.0$ and $E_{\max} = 100$.
We use the same ansatz shown in Fig.~\ref{fig-circuit} with 4 qubits and 4 layers.

Fig.~\ref{fig-rate} shows the dependence of the error $\epsilon(\vec{\theta}^*)$ in the optimized noisy cost function on the error rate $q$ in comparison with rough bounds (\ref{R1_BDs}), (\ref{b8n}) and (\ref{rough_LB}).
Here, the error probability of the two-qubit error and the readout error are $q$, and the error probability of the single-qubit error is $10^{-1}q$.
$E_1 - E_0$ is set to $50$.
All of our rough bounds well capture the scaling of the true error dependence on the error rate as similar to the Heisenberg spin chain.

On the other hand, Fig.~\ref{fig-gap} shows the dependence of the error $\epsilon(\vec{\theta}^*)$ in the optimized noisy cost function on the spectral gap $E_1 - E_0$ in comparison with rough bounds (\ref{R1_BDs}), (\ref{b8n}) and (\ref{rough_LB}).
Here the error probability of the single-qubit error is set to \cob{$10^{-3}$, and the error probability of the two-qubit error and the readout error is $10^{-2}$.}
\coc{According to Fig.~\ref{fig-gap}, up to the moderate size of the gap, all the bounds including extremely rough lower bound~(\ref{rough_LB}) actually work well. However, for large gaps, extremely rough lower bound (\ref{rough_LB}) breaks down and overestimates the error because the impact of the error in the rough approximation $G_{i_1,i_2,\cdots,i_k}(\vec{\theta}^*)\approx 1$ is emphasized by the large value of the gap $E_1 - E_0$.
Fig.~\ref{fig-gap} also implies that the error tends to increase as the gap gets large as implied by the lower bound.}


\section{Conclusion}
\label{jzbmlzhd}

\com{We have established analytic formulae for estimating the error in the cost function of VQAs due to Gaussian noise.
We can also apply our formulae to a wide class of stochastic noise including depolarizing noise via their equivalence with Gaussian noise.
The first main result Theorem \ref{thm1} gives the leading-order approximation of the error $\epsilon(\vec{\theta})$ in the cost function due to the noise.
The Hessian of the cost function as the coefficients of the noise effect implies a trade-off relation between the hardness of the optimization of the parameters and the noise resilience of the cost function.
We have derived an order estimation of the sufficient error probability to achieve a given precision based on this formula.
This estimation \coa{offers} stringently small error probability \coa{if no error mitigation is taken into account.
This is partially because, the estimation is nothing but a sufficient condition to achieve the given precision.}
On the other hand, the estimation of the necessary order of the error probability to achieve a given precision is provided \coc{based on the lower bound Eq.~(\ref{rough_LB}).}
Though this estimation actually gives a larger error probability, it is still stringent \coa{without any error mitigation.}

\coc{Theorem \ref{thm2} gives upper and lower bounds on the error $\epsilon_0(\vec{\theta})$ for approximating $E_0$.
Especially for a minimal point,}
these bounds show how the spectrum of the target operator and the geometry of the ansatz affect the sensitivity of the cost function to the noise.
The bounds also imply other trade-off relations between the hardness of the optimization and the noise resilience of the cost function attributed to the spectral gap property or the smallness of the Fubini-Study metric of the ansatz.
\coc{Although it is impractical to calculate the full expression Eq.~(\ref{main_ULbounds}) of the bounds,
we have also shown rough bounds which are easier to calculate.}
The numerical simulations of \coc{the VQE of the Heisenberg spin chain} and the toy model Hamiltonian have demonstrated the usefulness of our rough bounds.
These rough estimations may be utilized as a simple inspection to check the order of magnitude of the impact of the noise.

A highlight of \cob{the} applications of our formula is the proposal of a quantum error mitigation method shown in Sec.~\ref{jshqlzhb}.
The essence of this error mitigation method is the cancellation of the error based on the expansion of the error with respect to the fluctuations of the parameters including the virtual parameters.
An advantage of this method is that we only use the noisy estimation of the derivatives of the cost function to mitigate the error, and we do not need to change the noise strength as in the extrapolation method \cite{PhysRevX.7.021050,PhysRevLett.119.180509}, nor to sample various circuits as in the probabilistic error cancellation \cite{PhysRevLett.119.180509, PhysRevX.8.031027}.
Although the effectiveness of this method is still inconclusive since we have only an estimation of the sufficient order of the error probability for this method to work, there is a possibility of this method to be efficient.
It may also possible to improve this method by taking into account higher order expansions.
In a future work, further analysis will be done on this error mitigation method including the finiteness of the sampling and the comparison with other error mitigation methods.
To take into account the statistical error due to the finiteness of the sampling, the effect of the noise on the variance of the cost function should also be considered in future works.
}

\cod{\section{Code availability}
Code to reproduce the numerical simulations in this work is available at \cite{source.code}.
A python module to compute rough bounds (\ref{R1_BDs}), (\ref{b8n}) and (\ref{rough_LBc}) is also provided.
}

\begin{acknowledgments}
The authors would like to thank Yasunari Suzuki for very helpful comments.
This work is supported by MEXT Quantum Leap Flagship Program (MEXT QLEAP) Grant Number JPMXS0118067394 and JPMXS0120319794.
K. F. is supported by JSPS KAKENHI Grant No. 16H02211, JST ERATO JPMJER1601, and JST CREST JPMJCR1673.
W. M. wishes to thank Japan Society for the Promotion of Science (JSPS) KAKENHI No. 18K14181 and JST PRESTO No. JPMJPR191A.
We also acknowledge support from JST COI-NEXT Grant No. JPMJPF2014.
\end{acknowledgments}

\appendix
\clearpage
\begin{widetext}
\cod{\section{Table of notations}\label{App_notations}
\begin{table*}[h]
 \centering
 \caption{Summary of important notations.}
 \label{tab:symbols_parameters}
 \renewcommand{\arraystretch}{1.1}%
 \begin{tabular}{c c}
  \hline
  \multicolumn{2}{l }{Parameters, parametric circuit, and stochastic noise channels}\\
  \hline
  $\theta_i$ & optimizing parameters or parameters including virtual parameters if specified in the context\\
  $\xi_\nu$ & virtual parameters (see Sec.~\ref{ssec-correspond} for detail)\\
  $A_i$ & generating operator of $i$-th parametric gate satisfying $A_i^2 = I$\\
  $U_i(\theta_i)$ & parametric gate $\exp [-i \theta_i A_i / 2]$\\
  $U(\vec{\theta})$ & parametric circuit $\prod_{i=1}^{M} U_i(\theta_i) W_i$ with non-parametric gate $W_i$\\
  $\ket*{\phi(\vec{\theta})}$ & state $U(\vec{\theta})\ket{\phi}$ created by the parametric circuit from an input state $\ket{\phi}$\\
  $B_\nu$ & $\nu$-th stochastic error operator satisfying $B_\nu^2 = I$\\
  $p_\nu$ & error probability of $B_\nu$-error\\
  $\mathcal{E}_{B_{\nu},p_{\nu}}$ & stochastic noise channel $\mathcal{E}_{B_{\nu},p_{\nu}}(\rho) = (1 - p_{\nu}) \rho + p_{\nu} B_{\nu} \rho B_{\nu}$\\
  $M$ & number of optimizing parameters\\
  $M_{\mathrm{SNC}}$ & number of stochastic noise channels $\mathcal{E}_{B_{\nu},p_{\nu}}$\\
  $M_{\mathrm{tot}}$ & number $M + M_{\mathrm{SNC}}$ of total parameters including both optimizing and virtual parameters\\
  $M_{\mathrm{dom}}$ & number of parameters out of $M_{\mathrm{tot}}$ with dominant error probability\\
  $G_{i_1, i_2, \cdots, i_k} (\vec{\theta})$ & sensitivity of the state $\ket*{\phi(\vec{\theta})}$ to the $\pi$-shift of the parameters $\theta_{i_1}, \cdots, \theta_{i_k}$ defined in Eq.~(\ref{Giii})\\
  \hline
  \multicolumn{2}{l }{Target operators and cost function}\\
  \hline
  $H_l$, $\ket{\phi_l}$ $(l=1,2,\cdots,L)$ & $L$ target Hermitian operators and input states, respectively\\
  $E_{0, l}$ & minimum eigenvalue of $H_l$\\
  $E_{\max,l}$ & largest eigenvalue of $H_l$\\
  $H$, $\ket{\phi}$ & single target Hermitian operator and single input state, respectively\\
  $E_0$ & minimum eigenvalue of $H$\\
  $E_1$ & second smallest eigenvalue of $H$\\
  $E_{\max}$ & largest eigenvalue of $H$\\
  $C(\vec{\theta})$ & cost function $\sum_{l=1}^{L}\bra{\phi_l} U(\vec{\theta})^{\dagger} H_l U(\vec{\theta}) \ket*{\phi_l}$ to minimize\\
  $C(\vec{\theta}, \vec{\xi})$ & cost function with explicit dependence on the virtual parameters $\xi$\\
  $C_{\mathrm{noisy}}(\vec{\theta})$ & noisy cost function\\
  $\vec{\theta}^*$ & minimal point of the cost function\\
  \hline
  \multicolumn{2}{l }{Error and precision}\\
  \hline
  $\epsilon(\vec{\theta})$ & error due to the noise in the cost function evaluated at $\vec{\theta}$\\
$\epsilon_0(\vec{\theta})$ & deviation of the noisy cost function from the minimum eigenvalue $E_0$ at $\vec{\theta}$\\
$\delta(\vec{\theta})$ & noiseless precision $C(\vec{\theta}) - E_0$ of the cost function at $\vec{\theta}$ for the minimization task\\
  $\epsilon_{*}$ & given desired precision\\
  $R_{\mathrm{L}}(\vec{\theta})$ & terms related with the noiseless precision $\delta(\vec{\theta})$ in lower bound (\ref{low_bd1}) (defined in Eq.~(\ref{RL}))\\
  $R_{\mathrm{U}}(\vec{\theta})$ & terms related with the noiseless precision $\delta(\vec{\theta})$ in upper bound (\ref{up_bd1}) (defined in eq.~(\ref{RU}))\\
  \hline
  \multicolumn{2}{l }{Gaussian noise model}\\
  \hline
$\mathcal{G}_{A,\sigma}$ & Gaussian noise channel with the variance $\sigma^2$, with respect to $A$ defined in Eq.~(\ref{Gauss_ch_def})\\
  $f_{\sigma}$ & probability density function $e^{-\frac{\eta^2}{2\sigma^2}}/(\sqrt{2\pi}\sigma)$ of the zero-mean Gaussian distribution\\
  $\sigma_i^2$ & variance of the $i$-th parameter\\
  $\sigma_{\mathrm{SNC},\nu}^2$ & variance of the $\nu$-th virtual parameter associated with a stochastic noise channel\\
  \hline
  \multicolumn{2}{l }{Local depolarizing noise model}\\
  \hline
  $\mathcal{D}_{k, p}$ & $k$-qubit depolarizing channel with error probability $p$ defined in Eq.~(\ref{DP_def})\\
  $c_k$ & constant factor of the $k$-qubit error probability in the local depolarizing noise model\\
  $q$ & scaling of the error probability in the local depolarizing noise model\\
  $q_k$ & $k$-qubit error probability $(4^{k-1} - 4^{-1})c_k q$ in the local depolarizing noise model\\
  $M_{\mathrm{DP}}$ & number of the virtual parameters in the local depolarizing noise model\\
  $M_{\mathrm{DP, prop}}$ & number of the proper virtual parameters in the local depolarizing noise model\\
  \hline
 \end{tabular}
\end{table*}}
\end{widetext}

\cod{\section{Proofs of main results}\label{App_proofs_main}}
\cod{In this section, we present proofs of main theorems and propositions in the main text.}
\cod{\subsection{Proof of Proposition \ref{prop1}}\label{prop1_proof}
Here, we prove Proposition \ref{prop1}.
\begin{proof}
 We define a map
\begin{align}
 \mathcal{U}_{B_\nu,\Delta}(\rho):= e^{-i\frac{\Delta}{2}B_\nu}\rho e^{i\frac{\Delta}{2}B_\nu}.
\end{align}
Using the relation
\begin{align}
 e^{-i\frac{\Delta}{2}B_\nu} =  I \cos \frac{\Delta}{2} - i B_\nu \sin \frac{\Delta}{2},
\end{align}
we have
\begin{align}
 \mathcal{U}_{B_\nu,\Delta}(\rho) + \mathcal{U}_{B_\nu,-\Delta}(\rho)
= 2 \rho \cos^2 \frac{\Delta}{2} + 2 B_\nu \rho B_\nu \sin^2 \frac{\Delta}{2}.\label{UAjd}
\end{align}
From Eq.~(\ref{UAjd}), we obtain the equivalence between the Gaussian noise channel $\mathcal{G}_{B_\nu,\sigma_{\mathrm{SNC}, \nu}}$ with respect to $B_\nu$ with the variance
\begin{align}
 \sigma_{\mathrm{SNC}, \nu}^2 = - 2 \log (1 - 2 p_\nu)
\end{align}
and the given stochastic noise channel $\mathcal{E}_{B_\nu,p_\nu}$ as follows:
\begin{align}
&\mathcal{G}_{B_\nu,\sigma_{\mathrm{SNC}, \nu}}(\rho)\nonumber\\
=&\int_{-\infty}^{\infty} \mathcal{U}_{B_\nu,\Delta}(\rho) \frac{e^{-\frac{\Delta^2}{2\sigma_{\mathrm{SNC}, \nu}^2}}}{\sqrt{2\pi}\sigma_{\mathrm{SNC}, \nu}} d\Delta \nonumber\\
=& 2 \int_{0}^{\infty} \left(2 \rho \cos^2 \frac{\Delta}{2} + 2 B_\nu \rho B_\nu \sin^2 \frac{\Delta}{2}\right) \frac{e^{-\frac{\Delta^2}{2\sigma_{\mathrm{SNC}, \nu}^2}}}{\sqrt{2\pi}\sigma_{\mathrm{SNC}, \nu}} d\Delta\nonumber\\
=& (1 - p_\nu) \rho + p_\nu B_\nu \rho B_\nu\nonumber\\
=& \mathcal{E}_{B_\nu,p_\nu}(\rho).
\end{align}
\end{proof}}
\subsection{Proof of Theorem \ref{thm0}}\label{thm0_proof}
Here, we give a proof of Theorem \ref{thm0}.
\begin{proof}
 Let us introduce the multi-index notation for $\alpha \in \mathbb{N}^{M_{\mathrm{tot}}}$ and $\vec{\theta} \in \mathbb{R}^{M_{\mathrm{tot}}}$ as follows:
\begin{align}
 \vec{\theta}^{\alpha}:= \prod_{i=1}^{M_{\mathrm{tot}}} \theta_i^{\alpha_i}, \quad \alpha ! := \prod_{i=1}^{M_{\mathrm{tot}}} \alpha_i !, \quad |\alpha| := \sum_{i=1}^{M_{\mathrm{tot}}}\alpha_i.
\end{align}
 The partial derivatives of a function $f$ are denoted as
\begin{align}
 D^{\alpha} f := \frac{\partial^{|\alpha|}}{\partial \theta_1^{\alpha_1} \partial \theta_2^{\alpha_2} \cdots \partial \theta_M^{\alpha_M}} f.
\end{align}
By Taylor expanding the integrand $C(\vec{\theta} + \vec{\eta})$, we obtain the following expression from the definition of the noisy cost function \eqref{gen_Cnoisy}
\begin{align}
 &\tilde{C}(\vec{\theta}) \nonumber\\
=& C(\vec{\theta}) + \int \sum_{|\alpha| = 1}^{\infty} \frac{1}{\alpha !} D^{\alpha} C(\vec{\theta}) \vec{\eta}^{\alpha} d\mathcal{P}(\vec{\eta})
\nonumber\\
=& C(\vec{\theta}) + \frac{1}{2} \sum_{i=1}^{M_{\mathrm{tot}}} \frac{\partial^2}{\partial \theta_i^2} C(\vec{\theta}) \sigma_i^2 + \sum_{|\alpha| = 3}^{\infty} \frac{1}{\alpha !} D^{\alpha} C(\vec{\theta}) \int \vec{\eta}^{\alpha} d\mathcal{P}(\vec{\eta}),\label{ithelzgp}
\end{align}
where we denote $\prod_{i} d\mathcal{P}_i(\eta_i)$ by $d\mathcal{P}(\vec{\eta})$.

Because of $A_i^2 = 1$ the second derivatives read
\begin{align}
 \frac{\partial^2}{\partial \theta_i^2} C(\vec{\theta}) = \frac{1}{2} \left[C(\vec{\theta} + \pi \vec{e}_i) - C(\vec{\theta})\right], \label{even-d}
\end{align}
where $\vec{e}_i$ denotes the vector whose $i$-th component is $1$ and the other components are $0$.
Similar relation is used in Refs.~\cite{PhysRevResearch.2.013129,Huembeli_2021,PhysRevA.103.012405,Cerezo_2021}.
By recursively applying the relation \eqref{even-d},
it turns out that the derivatives $D^{2\alpha} C(\vec{\theta})$ have the form
\begin{align}
 D^{2\alpha} C(\vec{\theta}) = \frac{1}{2}\left[\frac{1}{2^{|\alpha|-1}}\sum_{i=1}^{2^{|\alpha|-1}}\left(C(\vec{\theta}_{i,1})-C(\vec{\theta}_{i,2})\right)\right]
\end{align}
with some parameters $\vec{\theta}_{i,1(2)}$.
Since $\sum_{l=1}^{L}E_{0,l} \leq C(\vec{\theta}) \leq \sum_{l=1}^{L}E_{\max,l}$ holds for any parameter $\vec{\theta}$, we obtain \cite{Kubler2020adaptiveoptimizer}
\begin{align}
 |D^{2\alpha} C(\vec{\theta})| \leq \frac{\sum_{l=1}^{L}(E_{\max,l} - E_{0,l})}{2}.\label{der-bd}
\end{align}
Then, applying Eq.~(\ref{der-bd}) to (\ref{ithelzgp}), we obtain
\begin{align}
&\left|\tilde{C}(\vec{\theta}) - C(\vec{\theta}) - \frac{1}{2} \sum_{i=1}^{M_{\mathrm{tot}}} \frac{\partial^2}{\partial \theta_i^2} C(\vec{\theta}) \sigma_i^2 \right|\nonumber\\
=&\left|\sum_{|\alpha| = 3}^{\infty} \frac{1}{\alpha !} D^{\alpha} C(\vec{\theta}) \int \vec{\eta}^{\alpha} d\mathcal{P}(\vec{\eta})\right|\nonumber\\
\leq& \frac{\sum_{l=1}^{L}(E_{\max,l} - E_{0,l})}{2} \sum_{|\alpha| = 3}^{\infty} \frac{1}{\alpha !} \left|\int\vec{\eta}^{\alpha} d\mathcal{P}(\vec{\eta})\right|\nonumber\\
=& \frac{\sum_{l=1}^{L}(E_{\max,l} - E_{0,l})}{2} \sum_{|\alpha| = 3}^{\infty} \frac{1}{\alpha !} \int\vec{\eta}^{\alpha} d\mathcal{P}(\vec{\eta}),
\label{TE}
\end{align}
where the last equality follows from the assumption of the nonnegativity of the moments $\int \eta_i^{\alpha} d \mathcal{P}_i(\eta_i) \geq 0$ (even moments are always positive).
Moreover, we have
\begin{align}
 &\sum_{|\alpha| = 3}^{\infty} \frac{1}{\alpha !} \int\vec{\eta}^{\alpha} d\mathcal{P}(\vec{\eta})\nonumber\\
=& \sum_{k = 3}^{\infty}\sum_{\sum_{i}\alpha_i = k} \frac{1}{\prod_i \alpha_i!}
\prod_i \int \eta_i^{\alpha_i} d\mathcal{P}_i(\eta_i)\nonumber\\
=& \prod_{i=1}^{M_{\mathrm{tot}}} \sum_{\alpha_i = 0}^{\infty} \frac{1}{\alpha_i !} \mu_i^{(\alpha_i)}
- \frac{1}{2}\sum_{i=1}^{M_{\mathrm{tot}}} \sigma_i^2 - 1 \nonumber\\
=& \prod_{i=1}^{M_{\mathrm{tot}}} g_i(1) - \frac{1}{2}\sum_{i=1}^{M_{\mathrm{tot}}} \sigma_i^2 - 1,
\end{align}
where $\mu_i^{(\alpha_i)}$ is the $\alpha_i$-th moment of $\mathcal{P}_i$ defined in the main text, and the Taylor expansion (\ref{taylor-mgf}) of the mgf is used to obtain the last equality.
Therefore, we obtain Eq.~(\ref{eq-thm1}).
\end{proof}

\cod{\subsection{Proof of Proposition \ref{prop2}}\label{sec_prop2}}
\cod{To estimate the second derivatives of $C(\vec{\theta}, \vec{\xi})$ with respect to the proper virtual parameters, let us consider a single variable function $C_{\nu}(\xi_\nu):= C(\vec{\theta}^*, \xi_\nu \vec{e}_\nu)$ of one virtual parameter $\xi_\nu$, where $\vec{e}_\nu$ is the $M_{\mathrm{SNC}}$-dimensional vector whose $\nu$-th component is $1$ and the others are $0$.
Since $C_{\nu}(\xi_\nu) = a_\nu \cos(\xi_\nu + b_\nu) + c_\nu$ holds \cite{nakanishi_sequential_2020}, where $a_\nu\geq 0$ and $b_\nu$ and $c_\nu$ are real numbers, we have
$\frac{\partial^2}{\partial \xi^{2}_\nu} C(\vec{\theta}^*) = \frac{d^2}{d \xi_\nu^2}C_\nu(0) = - a_\nu \cos(b_\nu)$.
Because the proper virtual parameters are not optimized, $\frac{\partial^2}{\partial \xi^{2}_\nu} C(\vec{\theta}^*)$ may be negative.
In this case, $a_\nu \cos(b_\nu) \geq 0$ holds.
Since $C((\vec{\theta}, \vec{\xi})) \geq E_0$ for any value of the parameters,
we have $C_{\nu}(\xi_\nu) \geq C_\nu(0) - \delta(\vec{\theta}^*)$, which implies
\begin{align}
 a_\nu \cos(\xi_\nu + b_\nu) \geq a_\nu \cos(b_\nu) - \delta(\vec{\theta}^*) \geq - \delta(\vec{\theta}^*).
\end{align}
Thus, we obtain $a_\nu \leq \delta(\vec{\theta}^*)$, and hence
\begin{align}
 \frac{\partial^2}{\partial \xi^{2}_\nu} C(\vec{\theta}^*) = -a_\nu \cos(b_\nu) \geq - a_\nu \geq - \delta(\vec{\theta}^*).\label{D2vp}
\end{align}
Because Eq.~(\ref{D2vp}) yields
\begin{align}
 & \frac{1}{2} \left[\sum_{i=1}^{M} \frac{\partial^2}{\partial \tilde{\theta}^{2}_i} C(\vec{\theta}^*) + \sum_{\nu=1}^{M_{\mathrm{DP}}} \frac{\partial^2}{\partial \xi^{2}_\nu} C(\vec{\theta}^*) c_{k_\nu}\right]q \nonumber\\
&+ O\left( \sum_{l=1}^{L}(E_{\max,l} - E_{0,l}) M^2 q^2 \right)
\nonumber\\
\geq & \frac{1}{2} \left[\sum_{i=1}^{M} \frac{\partial^2}{\partial \tilde{\theta}^{2}_i} C(\vec{\theta}^*) 
 - c \: M_{\mathrm{DP}} \: \delta(\vec{\theta}^*)\right] q \nonumber\\
&+ O\left( \sum_{l=1}^{L}(E_{\max,l} - E_{0,l}) M^2 q^2 \right),
\end{align}
the following inequality follows from Eq.~(\ref{err-est-q}):
\begin{align}
 \epsilon(\vec{\theta}^*)
\geq & \frac{1}{2} \left[\sum_{i=1}^{M} \frac{\partial^2}{\partial \tilde{\theta}^{2}_i} C(\vec{\theta}^*) 
 - c \: M_{\mathrm{DP}} \: \delta(\vec{\theta}^*)\right] q \nonumber\\
&+ O\left( \sum_{l=1}^{L}(E_{\max,l} - E_{0,l}) M^2 q^2 \right).\label{biofff}
\end{align}
\coc{Dividing by $q/2$ and then adding $c M_{\mathrm{DP}}\delta(\vec{\theta}^*) + O\left( \sum_{l=1}^{L}(E_{\max,l} - E_{0,l}) M^2 q \right)$ to both sides of inequality~(\ref{biofff}),
we obtain
\begin{align}
 & 2\frac{\epsilon(\vec{\theta}^*)}{q} + c M_{\mathrm{DP}}\delta(\vec{\theta}^*) + O\left(\sum_{l=1}^{L}(E_{\max,l} - E_{0,l}) M^2 q\right)
\nonumber\\
\geq &
\sum_{i=1}^{M} \frac{\partial^2}{\partial \tilde{\theta}^{2}_i} C(\vec{\theta}^*)
= \Tr \left|\left(\frac{\partial^2 C}{\partial \tilde{\theta}_i \partial \tilde{\theta}_j} (\vec{\theta}^*)\right)\right|.
\end{align}
The last equality follows from the positivity of the Hessian of the cost function at $\vec{\theta}^*$.}
}
\cod{\subsection{Proof of Theorem \ref{thm2}}\label{sec_thm_bounds}
We define the shifting map $\mathcal{S}_i$ which maps the cost function to the shifted one as $\mathcal{S}_i C(\vec{\theta}) = C(\vec{\theta} + \pi \vec{e}_i)$.
Then, the noisy cost function can be written as
\begin{align}
 C_{\mathrm{noisy}}(\vec{\theta}) = \prod_{i=1}^{M_{\mathrm{tot}}}\left(1-p_i + p_i \mathcal{S}_i\right)C(\vec{\theta}).\label{noisyCprod}
\end{align}
Based on Eq.~(\ref{noisyCprod}), we can expand the error $\epsilon_0(\vec{\theta})$ as
\begin{align}
 &\epsilon_0(\vec{\theta})\nonumber\\
=& C_{\mathrm{noisy}}(\vec{\theta}) - E_0\nonumber\\
=& \prod_{i=1}^{M_{\mathrm{tot}}}\left(1-p_i\right)(C(\vec{\theta}) - E_0) \nonumber\\
&+
\sum_{i=1}^{M_{\mathrm{tot}}} \prod_{j\neq i} (1-p_j) p_i (\mathcal{S}_i C(\vec{\theta}) - E_0)\nonumber\\
&+\sum_{i_1\neq i_2 =1}^{M_{\mathrm{tot}}} \prod_{j\neq i_1, i_2} (1-p_j) p_{i_1} p_{i_2} (\mathcal{S}_{i_1} \mathcal{S}_{i_2} C(\vec{\theta}) - E_0)
+\cdots \nonumber\\
&+ \left(\prod_{i=1}^{M_{\mathrm{tot}}} p_i\right) \left(\prod_{i=1}^{M_{\mathrm{tot}}}\mathcal{S}_i - E_0\right)\nonumber\\
=&
\prod_{i=1}^{M_{\mathrm{tot}}}\left(1-p_i\right)\delta(\vec{\theta})+
\prod_{j = 1}^{M_{\mathrm{tot}}} (1 - p_j) \nonumber\\
&\times\sum_{k=1}^{M_{\mathrm{tot}}}\sum_{i_1 \neq i_2 \neq \cdots \neq i_k} 
\frac{\prod_{l=1}^k p_{i_l}}{\prod_{l=1}^k (1-p_{i_l})} \left(\prod_{l=1}^k \mathcal{S}_{i_l}C(\vec{\theta}) - E_0\right).\label{e0expand}
\end{align}
Notably, the precision $\delta(\vec{\theta})$ of the noiseless cost function is separated from the error due to the noise in the above expansion.
Furthermore, the difference of the shifted cost function $\prod_{l=1}^k \mathcal{S}_{i_l}C(\vec{\theta})$ from $E_0$ can be estimated as follows:
\begin{lemma}\label{shift-ineq}
The following relation holds
 \begin{align}
  &(E_1 - E_0)G_{i_1, i_2 \cdots, i_k}(\vec{\theta}) - 2\sqrt{(E_1 - E_0) \delta(\vec{\theta})}\nonumber\\
\leq&
\prod_{l=1}^k \mathcal{S}_{i_l} C(\vec{\theta}) - E_0\nonumber\\
\leq&
(E_{\max} - E_0)G_{i_1, i_2 \cdots, i_k}(\vec{\theta}) + 2\frac{E_{\max} - E_0}{\sqrt{E_1 - E_0}} \sqrt{\delta(\vec{\theta})},
 \end{align}
\end{lemma}
Combining Eq.~(\ref{e0expand}) with Lemma \ref{shift-ineq}, we obtain Theorem \ref{thm2}.

To prove Lemma \ref{shift-ineq}, we show the following relation between the fidelity and the expectation.
\begin{lemma}\label{Lem-fid-e}
 For any state $\rho$,
\begin{align}
\frac{\tr \rho H - E_0}{E_{\max} - E_0}
\leq
 1 - \bra{\psi_0}\rho\ket{\psi_0}
\leq
\frac{\tr \rho H - E_0}{E_{1} - E_0} \label{fid-e}
\end{align}
holds.
\end{lemma}
\begin{proof}
 Since the eigenspace for $E_0$ is nondegenerate, we have a decomposition
 \begin{align}
  H = E_0 \ketbra{\psi_0} + B, \label{E0-B}
 \end{align}
where $B := H - E_0 \ketbra{\psi_0}$ has the support on the orthocomplement of $\ket{\psi_0}$, and only has eigenvalues larger than or equal to $E_1$.
Hence, 
\begin{align}
 E_1 (I - \ketbra{\psi_0}) \leq B \leq  E_{\max} (I - \ketbra{\psi_0}) \label{B-ineq}
\end{align}
holds.
Thus, we obtain
\begin{align}
 \tr \rho H = & E_0 \bra{\psi_0}\rho\ket{\psi_0} + \tr \rho B \nonumber\\
\geq & E_0 \bra{\psi_0}\rho\ket{\psi_0} + E_1 \tr \rho (I - \ketbra{\psi_0})\nonumber\\
= & (E_0 - E_1) \bra{\psi_0}\rho\ket{\psi_0} + E_1,
\end{align}
which yields the right-most inequality in \eqref{fid-e}.
Similarly, we also obtain
\begin{align}
 \tr \rho H = & E_0 \bra{\psi_0}\rho\ket{\psi_0} + \tr \rho B \nonumber\\
\leq & E_0 \bra{\psi_0}\rho\ket{\psi_0} + E_{\max} \tr \rho (I - \ketbra{\psi_0})\nonumber\\
= & (E_0 - E_{\max}) \bra{\psi_0}\rho\ket{\psi_0} + E_{\max},
\end{align}
which yields the left-most inequality in \eqref{fid-e}.
\end{proof}

Now, we prove Lemma \ref{shift-ineq}.
\begin{proof}[Proof of Lemma \ref{shift-ineq}]
 Applying decomposition \eqref{E0-B} and relation \eqref{B-ineq}, and defining $\ket*{\phi_{\bm{i}}(\vec{\theta})}:= \ket{\phi\left(\vec{\theta} + \pi\sum_{l=1}^k\vec{e}_{i_l}\right)}$ for $\bm{i}=(i_1,i_2,\cdots,i_k)$, we have
\begin{align}
 &\prod_{l=1}^k \mathcal{S}_{i_l} C(\vec{\theta}) \nonumber\\
=& E_0 \bra*{\phi_{\bm{i}}(\vec{\theta})}\ket*{\psi_0}\hspace{-3pt}\bra*{\psi_0}\ket*{\phi_{\bm{i}}(\vec{\theta})} + \bra*{\phi_{\bm{i}}(\vec{\theta})}B\ket*{\phi_{\bm{i}}(\vec{\theta})}\nonumber\\
\geq &
E_0 \bra*{\phi_{\bm{i}}(\vec{\theta})}\ket*{\psi_0}\hspace{-3pt}\bra*{\psi_0}\ket*{\phi_{\bm{i}}(\vec{\theta})}
+ E_1 \bra*{\phi_{\bm{i}}(\vec{\theta})}(I - \ketbra*{\psi_0})\ket*{\phi_{\bm{i}}(\vec{\theta})}\nonumber\\
=& E_1 - (E_1 - E_0) \bra*{\phi_{\bm{i}}(\vec{\theta})}\ket*{\psi_0}\hspace{-3pt}\bra*{\psi_0}\ket*{\phi_{\bm{i}}(\vec{\theta})}.
\end{align}
In the same way, we also have
\begin{align}
 \prod_{l=1}^k \mathcal{S}_{i_l} C(\vec{\theta})
\leq E_{\max} - (E_{\max} - E_0) \bra*{\phi_{\bm{i}}(\vec{\theta})}\ket*{\psi_0}\hspace{-3pt}\bra*{\psi_0}\ket*{\phi_{\bm{i}}(\vec{\theta})}.
\end{align}
Hence, we obtain
\begin{align}
&(E_1 - E_0)\left(1 -
\braket*{\phi_{\bm{i}}(\vec{\theta})}{\psi_0}\braket*{\psi_0}{\phi_{\bm{i}}(\vec{\theta})}\right)\nonumber\\
\leq &\prod_{l=1}^k \mathcal{S}_{i_l} C(\vec{\theta}) - E_0 \nonumber\\
\leq &
(E_{\max} - E_0)\left(1 -
\braket*{\phi_{\bm{i}}(\vec{\theta})}{\psi_0}\braket*{\psi_0}{\phi_{\bm{i}}(\vec{\theta})}\right).
\end{align}
The inner product $\bra*{\phi_{\bm{i}}(\vec{\theta})}\ket*{\psi_0}$ is related with the inner product $\bra*{\phi_{\bm{i}}(\vec{\theta})}\ket*{\phi(\vec{\theta})}$ as follows:
\begin{align}
 &\left|\braket*{\phi_{\bm{i}}(\vec{\theta})}{\psi_0}\braket*{\psi_0}{\phi_{\bm{i}}(\vec{\theta})} - \braket*{\phi_{\bm{i}}(\vec{\theta})}{\phi(\vec{\theta})}\braket*{\phi(\vec{\theta})}{\phi_{\bm{i}}(\vec{\theta})}\right|\nonumber\\
= & \left|\tr \left[\ketbra*{\phi_{\bm{i}}(\vec{\theta})} \left(\ketbra*{\psi_0} - \ketbra*{\phi(\vec{\theta})}\right)\right]\right|\nonumber\\
\leq & \left\|\ketbra*{\phi_{\bm{i}}(\vec{\theta})}\right\| \left\|\ketbra*{\psi_0} - \ketbra*{\phi(\vec{\theta})}\right\|_1\nonumber\\
=& 2\sqrt{1 - |\braket*{\phi(\vec{\theta})}{\psi_0}|^2}, \label{est-FI}
\end{align}
where $\|\circ\|_1$ denotes the trace norm.
Then, applying \eqref{est-FI}, we obtain
\begin{align}
&(E_1 - E_0)G_{i_1,i_2,\cdots,i_k}(\vec{\theta}) - 2(E_1 - E_0) \sqrt{1 - |\braket*{\phi(\vec{\theta})}{\psi_0}|^2}\nonumber\\
\leq &\prod_{l=1}^k \mathcal{S}_{i_l} C(\vec{\theta}) - E_0 \nonumber\\
\leq &
(E_{\max} - E_0)G_{i_1,i_2,\cdots,i_k}(\vec{\theta})\nonumber\\
&+ 2(E_{\max} - E_0) \sqrt{1 - |\braket*{\phi(\vec{\theta})}{\psi_0}|^2}.\label{key-ineq1}
\end{align}
Applying Lemma \ref{Lem-fid-e} with $\rho = \ketbra*{\phi(\vec{\theta})}$, we have
\begin{align}
1 - |\braket*{\phi(\vec{\theta})}{\psi_0}|^2
\leq \frac{C(\vec{\theta}) - E_0}{E_1 - E_0} = \frac{\delta(\vec{\theta})}{E_1 - E_0},
\end{align}
and hence, Eq.~\eqref{key-ineq1} further reads
\begin{align}
&(E_1 - E_0)G_{i_1,i_2,\cdots,i_k}(\vec{\theta}) - 2\sqrt{(E_1 - E_0)\delta(\vec{\theta})}\nonumber\\
\leq &\prod_{l=1}^k \mathcal{S}_{i_l} C(\vec{\theta}) - E_0 \nonumber\\
\leq &
(E_{\max} - E_0)G_{i_1,i_2,\cdots,i_k}(\vec{\theta}) + 2\frac{E_{\max} - E_0}{\sqrt{E_1 - E_0}} \sqrt{\delta(\vec{\theta})}.
\end{align}
\end{proof}}
\section{Proof of Lemma \ref{Lem-gauss-dep}}\label{proof_lem1}
In this section, we prove Lemma \ref{Lem-gauss-dep} for completeness.
We consider the vector space $\mathcal{V}_k$ consisting of the operators acting on $k$-qubit.
Then, the $k$-qubit Pauli operators $\{P_i | i=0,1,\cdots, 4^k - 1\}$ is a basis of $\mathcal{V}_k$.
It is convenient to consider the matrix representation of quantum channels with respect to this basis.
Since the Pauli channels $\mathcal{U}_{P_i}(\rho) = P_i \rho P_i$ are mutually commutative, they are simultaneously diagonalized in the Pauli basis $\{P_i | i=0,1,\cdots, 4^k - 1\}$.
Then, the calculation of the product $\prod_{i=1}^{4^k - 1} [(1-\tilde{p})\mathcal{I} + \tilde{p}\mathcal{U}_{P_i}]$ is reduced to the calculation of each diagonal component.
The $(j,j)$-component of $(1-\tilde{p})\mathcal{I} + \tilde{p}\mathcal{U}_{P_i}$ is $1-2\tilde{p}$ if $P_i$ anticommutes with $P_j$, otherwise $1$ (i.e. if $P_i$ commutes with $P_j$).
The number of the generators of the Pauli group which anticommute to each element $P_i$ is calculated as
\begin{align}
 2^k \sum_{r\leq k, r:\mathrm{odd}} \binom{k}{r} = 2^{k} 2^{k-1} = 2 \cdot 4^{k-1}.
\end{align}
Therefore, the matrix expression of $\prod_i [(1-\tilde{p})\mathcal{I} + \tilde{p}\mathcal{U}_{P_i}]$ in the Pauli basis is
\begin{align}
  \mathrm{diag} (1, (1-2\tilde{p})^{2\cdot 4^{k-1}}, \cdots, (1-2\tilde{p})^{2\cdot 4^{k-1}}).
\end{align}
On the other hand, the matrix expression of the $k$-qubit depolarizing channel is
\begin{align}
 \mathrm{diag} \left(1, 1-\frac{4^k}{4^k - 1}p, \cdots, 1-\frac{4^k}{4^k - 1}p \right).
\end{align}
Thus, $\prod_i [(1-\tilde{p})\mathcal{I} + \tilde{p}\mathcal{U}_{P_i}]$ is equal to the depolarizing channel with the error probability $p$ if
$\tilde{p}$ satisfies
\begin{align}
 2\log (1-2\tilde{p}) = \frac{1}{4^{k-1}} \log \left(1-\frac{4^k}{4^k - 1}p\right).
\end{align}
Hence, the variance of the corresponding Gaussian noise \eqref{Gauss-stoch} reads
\begin{align}
 \sigma_{\mathrm{DP}}^2(k) = - 2\log (1-2\tilde{p}) = - \frac{1}{4^{k-1}} \log \left(1-\frac{4^k}{4^k - 1}p\right).
\end{align}


%

\end{document}